\documentclass[journal]{IEEEtran}

\ifCLASSINFOpdf
   \usepackage[pdftex]{graphicx}
   \graphicspath{{../pdf/}{../jpeg/}}
   \DeclareGraphicsExtensions{.pdf,.jpeg,.png}
\else
\fi
\usepackage{algorithmicx}
\usepackage{algorithm}
\usepackage{algpseudocode}

\usepackage{adjustbox}
\usepackage{xcolor}

\usepackage{amsmath}
\usepackage{multirow}

  \usepackage[caption=false,font=footnotesize]{subfig}

\usepackage{stfloats}
\usepackage{url}

\begin{document}
\title{Joint Application Admission Control and Network Slicing in Virtual Sensor Networks}


\author{Carmen~Delgado,~Mar\'ia~Canales,~Jorge~Ort\'in,~Jos\'e Ram\'on~G\'allego,~Alessandro~Redondi, Sonda~Bousnina and~Matteo~Cesana% <-this % stops a space
	\thanks{C. Delgado, M. Canales and J.R. G\'allego are with the Arag\'on Institute of Engineering Research, Universidad de Zaragoza, Zaragoza, Spain, E-mail: jrgalleg@unizar.es.}% <-this % stops a space
	\thanks{J. Ort\'in is with Centro Universitario de la Defensa, Zaragoza, Spain, and with the Arag\'on Institute of Engineering Research, Universidad de Zaragoza, Zaragoza, Spain}% <-this % stops a space
	\thanks{A. Redondi, S. Bousnina and M. Cesana are with Dipartimento di Elettronica, Informazione e Bioingegneria, Politecnico di Milano, Milano, Italy.}}% <-this % stops a space
\markboth{}%
%\markboth{Preprint submitted to IEEE Internet of Things Journal}%
{Delgado \MakeLowercase{\textit{et al.}}: Bare Demo of IEEEtran.cls for IEEE Journals}



\maketitle

\begin{abstract}
We focus on the problem of managing a shared physical wireless sensor network where a single  network infrastructure provider leases the physical resources of the networks to application providers to run/deploy specific applications/services. In this scenario, we solve jointly the  problems of Application Admission Control (AAC), that is, whether to admit the application/service to the physical network, and wireless Sensor Network Slicing (SNS), that is, to allocate the required physical resources to the admitted applications in a transparent and effective way. We propose a mathematical programming framework to model the joint AAC-SNS problem which is then leveraged to design effective solution algorithms. The proposed framework is thoroughly evaluated on realistic wireless sensor networks infrastructures. \end{abstract}

\begin{IEEEkeywords}
Wireless Sensor Networks; Virtualization; Resource Allocation; Internet of Things; Optimization
\end{IEEEkeywords}

% \fi
\IEEEpeerreviewmaketitle

\section{Introduction}
\label{sec:introduction}

%WSNs impact and diffusion 
\IEEEPARstart{W}{ireless} Sensor Networks (WSNs) are one of the key enabling building blocks for the Internet of Things. Looking back at the evolution of WSNs, a clear trend can be observed moving from stand-alone, application-specific deployments to highly integrated wireless sensor systems used to support heterogeneous ecosystems of services and applications. In this context, the premium deployment domains for WSNs are nowadays Smart Cities, Smart Home and Buildings and Intelligent Transportation Systems, which are all characterized by the coexistence of sensor nodes with heterogeneous sensing, processing and communication capabilities, which all together support multiple applications and services. 

%WSNs application environment; heterogeneous ecosystems; smart city, smart building; very many applications that may interact at different levels;
The aforementioned trend calls for novel design good practices  to overcome the limits in flexibility, efficiency and manageability of vertical, task-oriented and domain-specific  WSNs. In this field, virtualization appears to be the most promising approach to reach this goal. Virtualization is already a consolidated reality at the very heart of cloud-based services in data centers and the Internet core, and is also gauging momentum in the domain of wireless mobile networks under the push of network ``softwareization''; as an example, one of the major innovation of the fifth generation (5G) of the mobile broadband systems under current standardization is network slicing, which allows allocating/partitioning the physical resources of the radio access network and the core network to multiple concurrent applications with heterogeneous requirements and constraints. 

%Reconfiguration is an issue. This leads to virtualization techniques. Mention slicing. 
Similar virtualization approaches have been recently proposed in the domain of WSNs to ease up reconfigurability and manageability of network resources, eventually opening up for novel business opportunities where the roles of WSN infrastructure provider and WSN service/application provider are decoupled. WSNs Virtualization includes all technologies to abstract the physical communication, sensing, and processing resources in a shared wireless sensor network to efficiently allocate them to multiple independent applications or clients. Generally speaking, the realization of virtual sensor networks requires technologies and solutions in different domains ranging from the node level, where the virtual sensor nodes must be able to support and run applications in a transparent way, up to the network level where effective platforms and solutions are required to manage and reconfigure on-the-fly the network resources.   

%Contributions: a shared sensor network operator; time varying applications, design of an optimal call admission control on the arriving applications. 
We focus here on the problem of managing a shared physical WSN over time; we look at the reference scenario where a single shared sensor network infrastructure provider owns an heterogeneous infrastructure which can be accessed by multiple application providers which issue requests to deploy specific applications/services. In this scenario, we solve jointly the  problems of application/service admission control (AAC), that is, whether to admit the application/service to the physical network, and wireless sensor network slicing (SNS), that is, to allocate the required physical resources to the admitted applications in a transparent and effective way.  We propose a mathematical programming framework to model the joint AAC-SNS problem, which is then leveraged to design effective solution algorithms. The proposed framework is thoroughly evaluated on realistic heterogeneous WSNs. 
 
 The manuscript is organized as follows: Section \ref{sec:related} gives an overview of the background and the most relevant related literature, further commenting on the main contributions of the present work. Section \ref{model} introduces the addressed problem and the reference scenario, whilst Section \ref{sec:optim} and Section \ref{sec:dynamic} introduce the optimization framework to solve the joint problem of application admission control and wireless sensor network slicing in the two cases where application arrival time and activation time are/are not known a priori, respectively. In Section \ref{sec:heuristic} we introduce a heuristic to solve the joint AAC-SNS problem whose performance is evaluated in Section \ref{performance}. Concluding remarks are finally reported in Section \ref{sec:conclusions}.

\section{Related Work}
\label{sec:related}
%paragraph on general virtualization in wired and wireless. Pointer to survey
Virtualization is a widely used technique in the management of cloud-based services in the core network of the Internet \cite{Xia2017}. Besides that, virtualization is also impacting the domain of wireless networks in general  and mobile radio networks in particular with the ongoing standardization of novel virtualization-aware features for the fifth generation (5G) of mobile radio network including network slicing and Cloud-RAN \cite{samdanis2017}. 
 
Only recently virtualization technologies have also been studied in the domain of WSNs with the primary goal to improve the flexibility, the manageability and the Return On Investment (ROI) of widespread and large WSNs deployments. Being the research field that recent, a common and shared terminology is still missing and the technical papers often use different wording for similar concepts; as an example, \emph{virtual sensor networks}, \emph{shared sensor networks}, \emph{federated sensor networks} and \emph{multi-application sensor networks} are often used almost interchangeably in the related literature. The interested reader may refer to the following surveys on the topic \cite{khan2015, Farias2016}. 
  
In this work, we will use the term \emph{virtual sensor network} to define a physical sensor network heterogeneous infrastructure which can be used to support multiple concurrent applications and services, and where the ownership of the physical infrastructure is decoupled from the ownership of the applications and services; along the same lines, we will refer to \emph{virtualization technologies} to define the different solutions and approaches to support and realize a \emph{virtual sensor network}.  

Virtualization technologies at different levels are needed to actually enable a virtual sensor network. One insightful classification of such technologies available in the literature distinguishes virtualization technologies at the node level and  at the network level \cite{khan2015}.  \emph{Node-level virtualization} encompasses the design of abstraction layers and primitives on the single sensor node to overcome the problem of application-platform dependency and/or code modularization; in this field, architectures based on virtual machines are proposed to enable virtualization and re-programmability. As an example, Mat\'e \cite{Levis2002}, ASVM \cite{Levis2005}, Melete \cite{Yu2006} and VMStar \cite{Koshy2005} are frameworks for building application-specific virtual machines over constrained sensor platforms.  Similarly ReLog \cite{Zhu2017132} proposes a systematic approach consisting of a programming language, a compiler, and a virtual machine to make application programs concise and easy to modify. 

\emph{Network-level virtualization} technologies include two main building blocks which are usually tightly coupled: (i) management platforms to support multiple application sharing a common physical infrastructure, and (ii) tools/algorithms to allocate the physical resources to the multiple applications. Representative examples of management platforms are SenSHare \cite{Leontiadis2012} and UMADE \cite{5465985}, which create multiple overlay sensor networks which are ``owned'' by different applications on top of a shared physical infrastructure. Along the same lines, Fok \emph{et al.} \cite{Fok2011}  and \cite{Li20141775} introduce middleware abstractions to support multiple applications in heterogeneous WSNs. A prototype of Software Defined Wireless Sensor Network is proposed in \cite{Huang2015IJDSN} where a centralized control plane dynamically manages communication routes in the network with the goal of augmenting the energy efficiency.  

As far as resource allocation in virtual sensor network is concerned, the authors of \cite{Xu2010} focus on environmental monitoring applications and propose a centralized optimization framework which allocates applications to sensor nodes with objectives to minimize the variance in sensor readings. The same authors address in a later work the case where the application assignment problem is no longer centralized but rather distributed by resorting to game-theoretic tools \cite{6195490}.  Ajmal \emph{et al.} \cite{Ajmal2014} propose a decision algorithm to dynamically ``admit'' applications to physical sensor network infrastructure. Activity time maximization of the physical infrastructure is targeted in \cite{6496703} and \cite{Zeng2015TC}, which focus on the problem of scheduling applications in shared sensor nodes. In our past work \cite{Delgado2016}, we study the problem of resource allocation in virtual sensor networks in a static case where the set of applications to be serviced is fixed and given. The research streamline on sensor mission assignment \cite{Johnson2010, Porta2014, Cheng2014, Edalat2016} also addresses a resource allocation problem in wireless sensor networks; namely, the problem is to jointly allocate the physical sensor resources to incoming applications while incorporating admission control policies. In this respect, the problem addressed in our work  bears similarities with sensor mission assignment; on the other side, our approach has the following distinctive additional features which make it more comprehensive with respect to the related work on sensor mission assignment: our framework considers networking-related aspects (e.g., routing, interference and network capacity), includes the possibility to re-configure the sensor network by moving applications which were previously deployed,  and further allows modelling situations where multiple applications can be concurrently deployed at a sensor node.

The present work naturally fits in the network-level virtualization class and further extends the related literature in this field by considering a dynamic case where applications/services are ``offered'' to the physical infrastructure provider over time. Our proposal provides a more flexible resource allocation, aligned with the ideas of virtualization and network slicing, and a more detailed modeling of application demands, sensor resources and networking aspects than previous sensor mission assignment works. To the authors' knowledge, our work is the first approach to model and analyze the joint problem of application admission control and physical resource allocation in virtual sensor networks.

\begin{table*}[t!]\footnotesize %\tiny%\scriptsize
	
	\renewcommand{\arraystretch}{1.3}
	\centering
	\caption{Set of parameters of the optimization framework.}
	\begin{tabular}{|c|l|}
		\hline
		Parameter & Definition \\
		\hline
		\hline
		
		\multicolumn{2}{|c|}{Sets, vectors, characteristics}\\
		\hline
		\hline
		
		$S = \left\{s_1,s_2, \dotsc , s_{\mid S \mid}\right\} $ & Set of sensor nodes; subscript index $i$ (or $h$) refers to sensor node $s_i$ (or $s_h$)  \\
		\hline
		$S'$ & Set of nodes that are not Sinks (a subset of $S$)\\
		\hline
		$A = \left\{a_1,a_2, \dotsc , a_{\mid A \mid}\right\}$ & Set of applications; subscript index $j$ refers to application $a_j$  \\
		\hline
		$W = \left\{w_1, w_2, \dotsc ,w_{\mid W \mid} \right\}$ & Set of test points; subscript index $k$ refers to test point $w_k$ \\
		\hline
		$W_j$ & Set of test points of application $j$ \\
		\hline
		$S_{jk}$ & Set of sensor nodes covering test point $k$ of application $j$ \\
		\hline
		$o_j = \left\{c_j,m_j,l_j\right\}$ & Requirement vector of application $j$ (source rate, memory, processing load)\\
		\hline
		$O_i = \left\{C_i, M_i, L_i, E_i\right\}$ & Resource vector of node $i$ (bandwidth, storage, proccessing power, energy)\\
		\hline
		$T = \left\{t_1, t_2,\dotsc ,t_{\mid T \mid}\right\}$ & Set of time instants; subscript index $n$ refers to time instant $t_n$\\
		\hline
		$\tau_j$ & Instant time in which application $j$ arrives\\
		\hline
		$\epsilon_j$ & Activity time of application $j$\\
		\hline
		$T_j$ & Subset of $T$ formed by the time instants that lay in the interval $\left[\tau_j, \tau_j + \epsilon_j \right)$  \\
		\hline
		$A_q = \left\{ a_j | \tau_q \in [\tau_q, \tau_q + \epsilon_q) \right\}$ & Set of applications that are running in the network at $\tau_q$, including the \textit{arriving} application\\
		\hline
		$Q = \left\{\tau_1, \tau_2,\dotsc ,\tau_{\mid A \mid}\right\}$ & Subset of $T$ containing the time instants of the arrival of the applications; subscript index $q$ refers \\ 
		& to time instant $\tau_q$\\
		\hline
		$R^{s}_{i}$ & Sensing range of node $i$\\
		\hline
		\hline

		\multicolumn{2}{|c|}{Coverage and application deployment}\\
		\hline
		\hline

		$z_j$ & Binary variable indicating if application $j$ is successfully deployed \\
		\hline
		$y_{ijkn}$ & Binary variable indicating if test point $k$ of application $j$ is deployed at sensor node $i$ at time instant $n$  \\
		\hline
		$x_{in}$ &  Binary variable indicating if sensor node $i$ is active in the network at time instant $n$\\
		\hline
		$h_{jkn}$ & Binary variable indicating if test point $k$ of set $T_j$ is sensed by a sensor node at time instant $n$\\
		\hline
		$y_{ijk}$ & Binary variable indicating if test point $k$ of application $j$ is deployed at sensor node $i$ \\
		\hline
		$x_{i}$ &  Binary variable indicating if sensor node $i$ is active in the network\\
		\hline
		$X_{i}$ & Constant equal to 1 if the node $i$ was active before the arrival of the new application, and 0 otherwise  \\
		\hline
		$Y_{ijk}$ & Constant equal to 1 if the test point $k$ of the application $j$ was being sensed in the node $i$ just before\\ 
		& the arrival of the new application, and 0 otherwhise \\
		\hline
		$N_{ij}$ & Maximum number of test points of application $j$ actually covered by sensor node $i$\\
		\hline
		$u_{in}$ & Binary variable:  $u_{in} = x_{in}x_{in-1}$\\
		\hline
		$v_{ijkn}$ & Binary variable:  $v_{ijkn} = y_{ijkn}y_{ijkn-1}$\\
		\hline

	\end{tabular}
	\label{table:param}
\end{table*}

\begin{table*}[t!]\scriptsize%\footnotesize \tiny%
	
	\renewcommand{\arraystretch}{1.3}
	\centering
	\caption{Set of parameters of the optimization framework.}
	\begin{tabular}{|c|l|}
		\hline
		Parameter & Definition \\
		\hline
		\hline
		
		\multicolumn{2}{|c|}{Propagation model}\\
		\hline
		\hline
		$P_{max}$ & Maximum transmission power\\
		\hline
		$R^{T}_{max}$ & Maximum transmission range with $P_{max}$\\
		\hline
		$R^{I}_{max}$ & Maximum interference range with $P_{max}$\\
		\hline
		$g_{ih}$ & Channel gain from transmitter $i$ to receiver $h$ in the directional link $\left(i,h\right)$\\
		\hline
		$d_{ih}$ & Distance from $i$ to $h$\\
		\hline
		$g_{0}$ & Constant dependent on antenna parameters\\
		\hline
		$\gamma$ & Path loss index\\
		\hline
		$\alpha$ & Receiver sensitivity\\
		\hline
		$\mu$ & Interference sensitivity\\
		\hline
		$p_{i}$ & Transmission power assigned to node $i$\\
		\hline
		$R^T{\left(p_i\right)}$ & Transmission range for node $i$ with transmission power $p_i$\\
		\hline
		$R^I{\left(p_i\right)}$ & Interference range for node $i$ with transmission power $p_i$\\
		
		\hline
		\hline

		\multicolumn{2}{|c|}{Routing}\\
		\hline
		\hline
		
		$f_{ihn}$ & Flow of data in bps transmitted from node $i$ to node $h$ at time instant $n$\\
		\hline
		$f_{ih}$ & Flow of data in bps transmitted from node $i$ to node $h$\\
		\hline
		$f_{ihj}$ & Flow of data for application $j$ in bps transmitted from node $i$ to node $h$\\
		\hline
		$r_{in}$ & Rate generated by node $i$ at time instant $n$  \\
		\hline
%		$K_{ih}$ & Constant higher than the maximum transmission rate of a node if nodes $i$ and $h$ are linked according to the DODAG routing tree; and $0$ otherwise \\
		$K_{ih}$ & If nodes $i$ and $h$ are linked according to the routing tree, $K_{ih}$ is a constant higher than the maximum transmission rate of a node. If not, $K_{ih}$ is $0$ \\
		\hline
		\hline
		\multicolumn{2}{|c|}{Bandwidth}\\
		\hline
		\hline
		$C_{ih}$ & Capacity of the link $\left(i,h\right)$\\
		\hline
		\hline
		\multicolumn{2}{|c|}{Energy}\\
		\hline
		\hline
		
		$\beta_1$ & $ \beta_1 = 50$ nJ/bit\\
		\hline
		$\beta_2$ & $\beta_2 = 0.0013 \text{pJ/bit/m}^4$\\
		\hline
		$\rho$ & $\rho = 50$ nJ/bit\\
		\hline
		$P^{t}_{in}$ & Power dissipation at the radio transmitter of node $i$ at time instant $n$\\
		\hline
		$P^{r}_{in}$ & Power dissipation at the radio receiver of node $i$ at time instant $n$\\
		\hline
		$P^{t}_{ij}$ & Power dissipation for the application $j$ at the radio transmitter of node $i$\\
		\hline
		$P^{r}_{ij}$ & Power dissipation for the application $j$ at the radio receiver of node $i$\\
		\hline
		$\varphi_i$ & Cost incurred every time a node $i$ is activated\\
		\hline
		$\delta_{ij}$ & Cost incurred in node $i$ due to the impact of receiving the bytecode of application $j$ \\
		\hline
		$\lambda_i$ & Variable indicating the residual energy that node $i$ would have once the activity time of the applications deployed in it or forwarded by it expires\\
		\hline
		$\lambda$ & Variable indicating the total residual energy of the network once the activity time of all the applications deployed in the network expires\\
		\hline

	\end{tabular}
	\label{table:param2}
\end{table*}

\section{Problem Statement and System Model}
\label{model}
	
The reference playground features one Shared Sensor Network Infrastructure Provider (SSN-IP) that owns a WSN composed of heterogeneous nodes in terms of processing, sensing and communication capabilities; the SSN-IP has full management control of the physical infrastructure and receives requests from Sensor Network Service Providers (SN-SP) to have their services, called \emph{applications} hereafter, deployed in the physical infrastructure. Each application request is characterized by an \emph{arrival time}, that is, the time the application request is issued, an \emph{activity time}, that is how long the application needs to be active in the network, and a \emph{set of requirements}, that is, the service level required by the application in terms of sensing area to be covered, required processing and storage capabilities of the node(s) hosting the application,  and required communication bandwidth  to deliver application data remotely across the physical sensor network. 
	
We set ourself in the perspective of the SSN-IP and address the problem on how to optimally manage the application requests coming from the SN-SPs to maximize our total revenue which is assumed to be proportional to the number of application requests which can be satisfied. In a nutshell, we introduce hereafter optimal and sub-optimal techniques to let the SSN-IP decide jointly when to admit/not admit an application request and how to re-configure consequently the current physical resource assignment to applications. 
	
Tables \ref{table:param} and \ref{table:param2} summarize the notation used through Sections \ref{model}-\ref{sec:dynamic}. Let $S = \left\{s_1,s_2, \dotsc , s_{|S|}\right\} $ be a set of sensor nodes, $A = \left\{a_1,a_2, \dotsc , a_{|A|} \right\}$ a set of applications which are to be deployed in the reference area, and~$W~=~\left\{w_1,w_2, \dotsc ,w_{|W|} \right\}$ a set of test points in the reference network scenario. These test points are physical locations where the application's sensing parameters must be measured (e.g., a test point can be a physical location where a temperature monitoring application needs to collect a temperature sample). To simplify notation, in the following we will use the subscript index $i$ (or $h$) to refer to a sensor node $s_i$ (or $s_h$), the subscript index $j$ to refer to an application $a_j$ and the subscript index $k$ to refer to a test point $w_k$.
	
Each application $j$ requires to sense a given set of test points $W_j \subseteq W$. Formally, the application $j$ has to be deployed in a subset of the sensor node set $S$ such that all the test points in $W_j$ are sensed. We consider that a test point is covered by a sensor node $i$ if it is within its sensing range, $R_i^{s}$. Thus, given a test point, a set of sensor nodes can \textit{cover} it (the test point can be in the sensing range of several nodes), but only one sensor node will \textit{sense} it.
	
Therefore, it is also convenient to introduce the set $S_{jk}$, defined as the set of sensor nodes which cover the test point $k$, with $k \in W_j$. In other words, if the application $j$ is deployed on any of the sensors in $S_{jk}$, then the target test point $k$ is sensed for this application. A necessary condition for an application $j$ to be successfully deployed is that all the test points in its target set $W_j$ must be sensed. 
	
To model the dynamics of the arrival and departure of applications, we assume that each application $j$ arrives and leaves the system at time instants $\tau_j$ and $\tau_j + \epsilon_j$, where $\epsilon_j$ is its activity time. With this, we can define an ordered set of time instants $T = \left\lbrace t_1, t_2, \dotsc, t_{|T|} \right\rbrace$, each of them corresponding to the moment at which the arrival or departure of an application happens. To simplify the notation, in the following we will use the subscript index $n$ to refer to the time instant $t_n$. Each application $j$ will be active in the network (if it is possible to deploy it) during a time interval that begins and ends at elements of $T$. We define $T_j$ as the subset of $T$ formed by the time instants that lay within the interval $[\tau_j, \tau_j + \epsilon_j)$, which corresponds to all the arrivals and departures of applications during the activity time of application $j$. 
	
Each application $j$ in $A$ is further characterized by a requirement vector $o_j = \left\{c_j,m_j,l_j\right\}$ which specifies the generated source rate [bit/s], memory [bits] and processing load [MIPS] consumed by the application when it is deployed on a sensor node. The requirement vector can be interpreted as the amount of resources needed to accomplish the specific tasks required by the application (e.g., acquire, process and transmit 10 temperature samples, or acquire process and transmit one JPEG image, etc.). Additionally, each sensor node $i$ in $S$ is characterized by a given resource vector $O_i = \left\{C_i, M_i, L_i, E_i\right\}$, which specifies its available bandwidth, storage capacity, processing power and energy. 
	
A protocol interference model with power control \cite{Shi13} is used to characterize the wireless communications among the sensor nodes. The maximum transmission power is $P_{\max}$. With this power, there are a maximum transmission range $R^T_{\max}$ and a maximum  interference range $R^I_{\max}$. Given a directional link between a pair of nodes $\left(i,h\right)$, the channel gain from transmitter $i$ to receiver $h$ is defined as $g_{ih} = g_0 d_{ih}^{-\gamma}$, being $d_{ih}$ the distance from $i$ to $h$, $\gamma$ the path loss index and $g_0$ a constant dependent on antenna parameters. In order for a transmission to be successful, the received power must exceed a threshold $\alpha$. Additionally, all the nodes under the interference range of a sensor node share the same transmission channel and therefore, the transmission time is shared among them.  If $p_i$ is the transmission power assigned to node $i$, a transmission towards $h$ is successful if $p_i g_{ih} > \alpha$. Thus, the transmission range for node $i$ with transmission power $p_i$ can be obtained as $R_i^T\left(p_i\right)=\left(p_i g_0/\alpha\right)^{1/\gamma}$. Similarly, the interference resulting from node $i$ with power $p_i$  is non-negligible only if it exceeds a certain threshold $\mu$. Then, the interference range is $R_i^I\left(p_i\right)~=~\left(p_i g_0/\mu\right)^{1/\gamma}$.
	
\section{General optimization framework}
\label{sec:optim}

We start off by casting the application admission control with wireless sensor network slicing (AAC-SNS) problem in the ideal case where  the activity time and the arrival time of each application are known a priori. This leads to a performance benchmark which will be then used to evaluate the algorithms and solutions in the real case where application arrivals cannot be predicted a priori.  In this reference scenario, we target the maximization of the overall serviced applications, subject to node-related and network-related constraints. Namely, this problem is subject to coverage constraints (the set of test points of each application must be sensed) and to application requirements (each application should be assigned enough processing and storage resources during its whole activity time to operate successfully). In addition, due to the multi-hop nature of WSNs, routing and link capacity constraints must be considered when the data generated by the applications have to be delivered remotely. 

Let $z_j$ be a binary variable indicating if application $j$ is successfully deployed in the network. Let $y_{ijkn}$ be a binary variable indicating if test point $k$ of application $j$ is deployed at sensor node $i$ at time instant $n$. Let $x_{in}$ be a binary variable indicating if sensor node $i$ is active in the network at time instant $n$. Let $h_{jkn}$ be a binary variable which indicates if test point $k$ of application $j$ is sensed at time instant $n$. The objective function aims to maximize the total number of deployed applications:

\begin{equation}
\label{eq:maximiza}
\max \sum_{j \in A}   z_j 
\end{equation}

\subsection{Constraints on coverage and on resources of the sensors}
\label{coverage-const}
	
Constraints (\ref{eq:const2})-(\ref{eq:const6}) require that all the applications which are actually deployed do fulfill the coverage constraints, that is, they sense all the required test points during their corresponding activity time. Specifically, Eq. (\ref{eq:const2}) indicates that a test point $k$ of an application $j$ is sensed at time instant $n$ if the application $j$ is deployed at a sensor node $i$ belonging to $S_{jk}$ at that instant. If so, it ensures that it is only sensed by sensor node $i$. Eq. (\ref{eq:const3}) ensures that if a sensor node $i$ does not cover a test point $k$ of an application $j$, then it can not sense that test point. Eq. (\ref{eq:const4}) guarantees that the test points of each application are only tested during the required time interval of the application $T_j$. Depending on the application, it can be possible that the same sensor node can sense several of its test points (e.g., visual applications). If we define $N_{ij}$ as the maximum number of test points of the same application $j$ that a sensor $i$ is able to sense, Eq. (\ref{eq:const5}) guarantees that this threshold is not exceeded at any time instant. Eq. (\ref{eq:const6}) indicates that an application $j$ is successfully deployed, i.e., $z_j = 1$, if and only if all its test points are sensed during the activity time of the application ($h_{jkn} = 1, \forall k \in W_j, \forall n \in T_j$). On the other hand, if the application is not successfully deployed, i.e., if $z_j = 0$, the constraint forces none of its test points to be sensed at all ($h_{jkn}=0, \forall k \in W_j, \forall n \in T_j$ and consequently according to Eq. (\ref{eq:const2}), $y_{ijkn}=0, \forall k \in W_j, \forall i \in S_{jk}, \forall n \in T_j$). This guarantees that if the application cannot be deployed, resources are not wasted.

\begin{equation}
	\sum_{i \in S_{jk}} y_{ijkn} = h_{jkn}  \qquad  \forall j \in A, \forall k \in W_j,  \forall n \in T \label{eq:const2}
\end{equation}
\begin{equation}
	y_{ijkn} = 0  \qquad  \forall i \notin S_{jk}, \forall j \in A, \forall k \in W_j,  \forall n \in T \label{eq:const3} 
\end{equation}
\begin{equation}
	y_{ijkn} = 0  \qquad  \forall i \in S_{jk}, \forall j \in A, \forall k \in W_j,  \forall n \notin T_j \label{eq:const4} 
\end{equation}
\begin{equation}
	\sum_{k \in W_j} y_{ijkn} \leq N_{ij}  \qquad \forall i \in S,  \forall j \in A, \forall n \in T_j  \label{eq:const5} 
\end{equation}
\begin{equation}
	z_j =  \frac{1}{\left| W_j \right| \left| T_j \right|} \sum_{n \in T_j} \sum_{k \in W_j} h_{jkn}  \qquad   \forall j \in A \label{eq:const6} 
\end{equation}
	
	Constraints (\ref{eq:const7}) and (\ref{eq:const8}) are budget-type constraints for the available storage and processing load of the sensor nodes. They must be ensured for all the nodes at any time instant $n \in T$.
	
	\begin{equation}
	\sum_{j \in A} \sum_{ k \in W_j} m_j y_{ijkn} \leq M_i  \qquad   \forall i \in S,  \forall n \in T \label{eq:const7}
	\end{equation}
	\begin{equation}
	\sum_{j \in A} \sum_{ k \in W_j} l_j y_{ijkn} \leq L_i  \qquad   \forall i \in S,  \forall n \in T \label{eq:const8}
	\end{equation}

\subsection{Routing constraints}
\label{routing-const}

Deployed applications require that the information generated locally is delivered remotely to collection points (sink nodes) through multihop paths. Note that these sensor nodes may run deployed applications or not. By resorting to a fluid model, it should be ensured that all the data produced by the sensors running applications at any time is received by the sink nodes. This fact can be conveniently expressed using the following constraints:

\begin{equation}
\label{eq:const9b}
r_{in} = \sum_{j \in A}\sum_{ k \in W_j } c_j y_{ijkn}  \quad \forall i \in S, \forall n \in T
\end{equation}
\begin{equation}
\label{eq:const9}
\sum_{\substack{ h \in S \\ i \neq h }}f_{hin} - \sum_{\substack{h \in S \\ h \neq i} }f_{ihn} + r_{in} = 0   \quad \forall i \in S', \forall n \in T
\end{equation}
\begin{equation}
\label{eq:const10}
\sum_{j \in A}  c_j \sum_{k \in W_j} h_{jkn} = \sum_{ h \in S \setminus S' } \left(\sum_{\substack{i \in S \\ i \neq h}} f_{ihn} + r_{hn} \right)  \quad \forall n \in T
\end{equation}

\noindent where $S'$ is the set of nodes that are not sinks (a subset of $S$), $r_{in}$ is a variable indicating the data generated by node $i$ at time instant $n$ and $f_{ihn}$ is a variable representing the flow of data in bps transmitted from node $i$ to node $h$ at time instant $n$. Constraints (\ref{eq:const9b}) and (\ref{eq:const10}) enforce flow conservation at sensor nodes: the incoming flow rate into a node $i$ plus the data generated by itself must be equal to the outcoming flow rate. When a node $i$ runs application $j$ to monitor its test point $k$, it generates traffic at a rate $c_j$ bps.

Constraints (\ref{eq:const10}) impose that the amount of data generated in the network at any time instant is equal to the amount of data collected by the set of sinks. The left term represents the total data rate generated in the network: for each active application $j$, there may be different sensor nodes sensing the corresponding $|W_j|$ test points and therefore each one generating $c_j$ bps. The right term represents the overall traffic received by the set of sink nodes plus the rate generated by themselves in case they are also running applications. This equality, together with the flow conservation in constraints (\ref{eq:const9}), imposes that all the traffic generated in the network is finally collected by the set of sinks (delivered by other nodes or generated by the sinks themselves).

The following constraints set enforces that if a sensor node is either running an application or receiving data at time instant $n$, then it must be active in the network and inactive otherwise:

\begin{equation}
\label{eq:const11}
x_{in} \leq  \sum_{\substack{h \in S \\ h \neq i}}f_{hin} + r_{in} \leq K x_{in}  \qquad   \forall i \in S, \forall n \in T
\end{equation}

\noindent where $K$ is a constant high enough (higher than the maximum transmission rate of a node). 

Finally, in WSNs routes from each sensor node to a sink node follow typically a single path, such as the Destination Oriented Directed Acyclic Graph (DODAG) of RPL \cite{rfc6550}. To model this behaviour, we build a DODAG for each sink using the number of hops as a metric (i.e., when there are several sinks, each node belongs to the DODAG that reaches a sink with the minimum number of hops). The following set of constraints forces all the traffic to be forwarded only through the links that belong to the predefined routes defined in the DODAGs:

\begin{equation}
\label{eq:const14}
f_{ihn} \leq K_{ih}    \qquad   \forall i, h \in S, \forall n \in T  
\end{equation}

\noindent where for each node $i$, $K_{ih}$ is $0$ for all nodes $h$ but for the father node in the routing tree towards the sink. In that case, $K_{ih}$ is a constant higher than the maximum transmission rate of a node. 

\subsection{Bandwidth constraints}
\label{bandwidth-const}

The available bandwidth in the network is bounded and must be shared among sensor nodes. We assume that a fair medium access control scheme orchestrates the access to the shared medium. Given a directional link between a pair of nodes $\left(i,h\right)$, let the capacity of the link be defined as $C_{ih}~=~\min\left(C_i, C_h\right)$. This aims to model that the transmission rate is limited by the most restrictive node in the link. Transmissions of other links where $i$ or $h$ are either transmitter or receiver cannot be simultaneously active with $\left(i,h\right)$ (note that some combinations are not possible in this particular case due to routing constraints, i.e., another link with $i$ as a transmitter). 

According to the considered protocol interference model, the interfering links for link $\left(i,h\right)$ are those whose receiver is within the interference range of node $i$ or the links where $h$ is within the interference range of its transmitter. Although none of these links can be simultaneously active with $\left(i,h\right)$, some of them (depending on their relative positions) could be simultaneously active with each other. Therefore, for each link in the network $\left(i,h\right)$ it must be ensured that the fraction of time used by the link plus all its interferences at time instant $n$ is less or equal to 1:

\begin{multline}
\label{eq:constf13}
\frac{f_{ihn}}{C_{ih}} +  \sum_{g \in S} \frac{f_{gin}}{C_{gi}} + \sum_{\substack{g \in S \\ g\neq i}} \frac{f_{hgn}}{C_{hg}} + \sum_{\substack{g \in S \\ g\neq i}} \frac{f_{ghn}}{C_{gh}}  \\ + \sum_{\substack {g,t \in S \\ d_{it}<R_i^I(p_i) }} \!\!\!\!\!\! \frac{f_{gtn}}{C_{gt}} + \!\!\!\!\!\! \sum_{\substack {g,t \in S \\ d_{gh}<R_g^I(p_g) }} \!\!\!\!\!\! \frac{f_{gtn}}{C_{gt}} \leq 1 \quad   \forall i,h \in S, \forall n \in T  
\end{multline}

Constraints (\ref{eq:constf13}) are the equivalent budget-type constraints for the available wireless capacity to the storage and processing load constraints given in (\ref{eq:const7}) and (\ref{eq:const8}). 

\subsection{Energy constraints}
\label{energy-const}

Finally, energy constraints are included to ensure that the application deployment pattern does not exceed the energy budget of the network. Typically, energy consumption due to wireless communication (i.e., transmitting and receiving) has been considered the dominant factor in power consumption for WSNs \cite{Akyildiz02}. While this is the case for traditional scalar applications, where processing is limited to simple operations, in multimedia applications the energy required to process data can not be neglected \cite{Redondi12}.

Regarding wireless transceiver, the power dissipation at the radio transmitter $P^{t}_{in}$ or at the radio receiver $P^{r}_{in}$ of each node $i$ at time instant $n$ can be modelled as \cite{Hou08}:
\begin{equation}
\label{eq:powertxall}
P^{t}_{in} = \sum_{h \in S, h \neq i} \left(\beta_1 + \beta_2  d^{\gamma}_{ih} \right)  f_{ihn}   \qquad   \forall i \in S, \forall n \in T 
\end{equation}
\begin{equation}
\label{eq:powerrxall}
P^{r}_{in} = \rho  \sum_{h \in S, h \neq i} f_{hin}   \qquad   \forall i \in S, \forall n \in T
\end{equation}

Typical values for $\beta_1$, $\beta_2$ and $\rho$ are $ \beta_1 =\rho = 50$ nJ/bit and  $\beta_2 = 0.0013 \text{pJ/bit/m}^4$,  with $\gamma = 4$ the path loss index. 

The estimation of the power dissipation due to the processing load is not so straightforward, since it depends on several factors such as the hardware architecture of the nodes or the specific implementation of the algorithm for each application. In the energy constraints set in (\ref{eq:const17a}), this power dissipation is left as a function $f$ of the processing loads $l_j$ of the applications. In Section~\ref{performance}, further details about the specific evaluated multimedia applications are given. In addition, we assume that the sinks do not have energy constraints since they can be plugged directly into the grid.

\begin{equation}
\label{eq:const17a}
\sum_{n \in T}{\Delta t_n \!\! \left( \!\! P^{t}_{in} \!  + \! P^{r}_{in} \! +  \! f \! \left(\sum_{j \in A} \sum_{ k \in W_j} y_{ijkn}  l_j\right) \!\! \right)} \! \leq E_i  \!\!\!\! \qquad   \forall i \in S'  
\end{equation}

To introduce the effect of the activation and deactivation of sensor nodes and the migration of active applications between sensor nodes of the network, we consider that both events decreases the energy of the affected nodes (i.e. the energy of a node decreases when it is booted-up or when it receives an application). In other words, there is a cost $\varphi_i$ related to the amount of energy that the node $i$ needs to wake up from the sleep mode. Besides, it is allowed moving applications from one active node to another as long as all the restrictions described previously are fulfilled. Nevertheless, it is assumed that moving an application $j$ has a cost $\delta_{ij}$ due to the impact of moving the application code to the new node $i$. This modifies the previous set of restrictions as follows:

\begin{multline}
\label{eq:const17}
\sum_{n \in T}{\Delta t_n \left(P^{t}_{in} + P^{r}_{in} + f\left(\sum_{j \in A} \sum_{ k \in W_j} y_{ijkn}  l_j\right)\right)} \\ + \varphi_i \sum_{n \in T} (x_{in} - x_{in-1})x_{in} \\ + \sum_{j \in A} \delta_{ij}\sum_{ k \in W_j} \sum_{n \in T} \left(y_{ijkn} - y_{ijkn-1}\right) y_{ijkn} \leq E_i
% \quad   \forall i \in S'  
\end{multline}

Considering that $x_{in}$ and $y_{ijkn}$ are binary variables and defining the binary variables $u_{in} = x_{in}x_{in-1}$ and $v_{ijkn}~=~y_{ijkn}y_{ijkn-1}$, the previous expression can be re-formulated as:
\begin{multline}
\label{eq:const18}
\sum_{n \in T}{\Delta t_n \left(P^{t}_{in} + P^{r}_{in} + f\left(\sum_{j \in A} \sum_{ k \in W_j} y_{ijkn}  l_j\right)\right)} \\ + \varphi_i \sum_{n \in T} (x_{in} - u_{in}) \\ + \sum_{j \in A} \delta_{ij}  \sum_{ k \in W_j} \sum_{n \in T} \left(y_{ijkn} - v_{ijkn} \right)  \leq E_i \qquad   \forall i \in S' 
\end{multline}

\noindent with the variables $u_{in}$ and $v_{ijkn}$ fulfilling the following set of restrictions:

\begin{equation}
u_{in} \leq x_{in} \qquad  \forall i \in S, \forall n \in T \label{eq:const113}
\end{equation}
\begin{equation}
u_{in} \leq x_{in-1} \qquad  \forall i \in S, \forall n \in T \label{eq:const113b}
\end{equation}
\begin{equation}
u_{in} \geq x_{in} + x_{in-1} - 1 \qquad  \forall i \in S, \forall n \in T \label{eq:const113c} 
\end{equation}
\begin{equation}
v_{ijkn} \leq y_{ijkn}  \qquad  \forall i \in S, \forall j \in A, \forall k \in W_j,  \forall n \in T \label{eq:const114a}
\end{equation}
\begin{equation}
v_{ijkn} \leq y_{ijkn-1}  \qquad  \forall i \in S, \forall j \in A, \forall k \in W_j,  \forall n \in T \label{eq:const114b} 
\end{equation}
\begin{multline}
v_{ijkn} \geq y_{ijkn} + y_{ijkn-1} - 1 \\ 
\qquad  \forall i \in S, \forall j \in A, \forall k \in W_j,  \forall n \in T \label{eq:const114c}
\end{multline}

\section{Optimization model for dynamic resource allocation}
\label{sec:dynamic}

The general optimization framework set in the previous section assumes that the arrival and departure times of each application are known in advance. While this approach provides a benchmark for the best achievable performance, its application to real scenarios is limited to very specific situations where this information could be available.

Thus, to encompass a more realistic scenario, in this section we focus on the problem of dynamically addressing the deployment of new applications as they arrive to the network. Specifically, the problem can be described as follows: every time a new application arrives to the reference area, it will be deployed as long as the rest of applications that are running at that moment are not dropped out of the system. If this is not possible, the new application is discarded and the system remains unaltered so that the previous applications can be served. If it is possible, the applications should be deployed in such a way that the likelihood of deploying new applications in the future is maximized.

As one of the most limiting factors to deploy new applications is the remaining energy of the nodes (it will be shown in Section \ref{performance}), three approaches are considered to select the way that applications are deployed in the network: (i) to maximize the global energy of the network, (ii) to maximize the energy of the node with the lowest energy, and (iii) a weighted sum of the first and the second approaches.

In the following, we define formally the optimization problem that must be solved every time a new application $q \in A$ arrives to the system. Let $Q = \left\{\tau_1,\tau_2, \dotsc , \tau_{|A|} \right\}$ be the subset of $T$ that contains all the arrival time instants. For every $\tau_q$, we define the subset $A_q \subseteq A$ as the set of applications that are running in the network at time instant $\tau_q$, \emph{including} the arriving application, i.e.  $A_q = \left\{ a_j | \tau_q \in [\tau_j, \tau_j + \epsilon_j) \right\}$. 

As the optimization problem that must be solved when the application $q$ arrives only involves those applications that are active in the network at time instant $\tau_q$ (i.e., those is $A_q$), we can remove the dependency on $n$ from the variables $y_{ijkn}$ and $x_{in}$  and redefine them as follows: $y_{ijk}$ is a binary variable indicating if application $j$ is deployed at a sensor node $i$ covering test point $k$, and $x_{i}$ is a binary variable indicating if sensor node $i$ is active in the network.

The following sets of restrictions force all the applications in $A_q$ to be deployed. It is worth noting that now the problem may be infeasible. If so, the system is left as it is to ensure that the current applications remain in the network and the new application is not deployed. Many of the following constraints are similar to the ones presented previously but without the subscript index $n$ and restricted to the subset $A_q$ instead of the set $A$.

\subsection{Constraints on coverage and on resources of the sensors}

Constraints (\ref{eq:constd211}) require that all the applications in $A_q$ sense all their required test points. To do so, it forces that for every test point $k$ of an application $j$, the application is deployed at a sensor node $i$ that can cover that test point $k$. Eq. (\ref{eq:constd212}) and (\ref{eq:constd213}) are equivalent to the constraints presented in (\ref{eq:const3}) and (\ref{eq:const5}) respectively.
\begin{equation}
\sum_{i \in S_{jk}} y_{ijk} = 1  \qquad  \forall j \in A_q, \forall k \in W_j  \label{eq:constd211}
\end{equation}
\begin{equation}
y_{ijk} = 0  \qquad  \forall j \in A_q, \forall k \in W_j, \forall i \notin S_{jk} \label{eq:constd212} 
\end{equation}
\begin{equation}
\sum_{k \in W_j} y_{ijk} \leq N_{ij}  \qquad \forall i \in S,  \forall j \in A_q  \label{eq:constd213} 
\end{equation}

Constraints (\ref{eq:constd214}) and (\ref{eq:constd215}) are budget-type constraints, equivalent to the equations in (\ref{eq:const7}) and (\ref{eq:const8}).
\begin{equation}
\sum_{j \in A_q} \sum_{ k \in W_j} m_j y_{ijk} \leq M_i  \qquad   \forall i \in S  \label{eq:constd214}
\end{equation}
\begin{equation}
\sum_{j \in A_q} \sum_{ k \in W_j} l_j y_{ijk} \leq L_i  \qquad   \forall i \in S  \label{eq:constd215}
\end{equation}

\subsection{Routing and bandwidth constraints}

Let $f_{ihj}$ be the flow of data of application $j$ in bps transmitted from node $i$ to node $h$ and $f_{ih}$ the flow of data in bps transmitted from node $i$ to node $h$. While $f_{ih}$ is equivalent to the already shown $f_{ihn}$, $f_{ihj}$ is needed to describe the corresponding power dissipation at radio transmitter and radio receiver as it will be shown in Section~\ref{dyn-energy}.

Constraints (\ref{eq:const9bii})-(\ref{eq:constd23}) are equivalent to constraints (\ref{eq:const9b})-(\ref{eq:constf13}). Now, in Eq. (\ref{eq:constd223}), since the problem forces all the applications to be active, the left term is the sum of all the  possible data rate generated in the network by all the applications in $A_q$ and it does not depend on any variable.

\begin{equation}
\label{eq:const9bii}
r_{ij} = \sum_{ k \in W_j } c_j y_{ijk}  \quad \forall j \in A_q, \forall i \in S
\end{equation}
\begin{equation}
f_{ih} = \sum_{j \in A_q} f_{ihj}  \quad    \forall i, h \in S 
\label{eq:constd221} 
\end{equation}
\begin{equation}
\sum_{\substack{ h \in S \\ i \neq h }}f_{hij} - \sum_{\substack{h \in S \\ h \neq i} }f_{ihj} + r_{ij} \qquad \forall j \in A_q, \forall i \in S'
\label{eq:constd222} 
\end{equation}
\begin{equation}
\sum_{j \in A_q} |W_j|  c_j = \sum_{ h \in S \setminus S' } \left(\sum_{\substack{i \in S \\ i \neq h}} f_{ih}   + \sum_{j \in A_q} r_{hj} \right)
\label{eq:constd223} 
\end{equation}
\begin{equation}
x_i \leq \sum_{\substack{h \in S \\ h \neq i}}f_{hi} + \sum_{j \in A_q}r_{ij} \leq K x_i  \qquad   \forall i \in S  
\label{eq:constd224} 
\end{equation}
\begin{equation}
f_{ih } \leq K r_{ih}    \qquad   \forall i,h \in S  \label{eq:constd227} 
\end{equation}
\begin{multline}
\frac{f_{ih}}{C_{ih}} + \sum_{g \in S} \frac{f_{gi}}{C_{gi}} + \sum_{\substack{g \in S \\ g\neq i}} \frac{f_{hg}}{C_{hg}} +  \sum_{\substack{g \in S \\ g\neq i}} \frac{f_{gh}}{C_{gh}}  \\ + \!\!\!\!\!\! \sum_{\substack {g,t \in S \\ d_{it}<R_i^I(p_i) }} \!\!\!\!\!\! \frac{f_{gt}}{C_{gt}} + \!\!\!\!\!\! \sum_{\substack {g,t \in S \\ d_{gh}<R_g^I(p_g) }} \!\!\!\!\!\! \frac{f_{gt}}{C_{gt}} \leq 1  \qquad   \forall i,h \in S   \label{eq:constd23} 
\end{multline}

\subsection{Energy constraints}
\label{dyn-energy}

The power dissipation for the application $j$ at the radio transmitter $P^{t}_{ij}$ or at the radio receiver $P^{r}_{ij}$ of each node $i$ can be modeled as: 

\begin{equation}
P^{t}_{ij} = \sum_{h \in S, h \neq i} \left(\beta_1 + \beta_2  d^{\gamma}_{ih} \right)  f_{ihj}   \qquad   \forall i \in S, \forall j \in A_q 
\label{eq:powertxalld}
\end{equation}
\begin{equation}
P^{r}_{ij} = \rho  \sum_{h \in S, h \neq i} f_{hij}   \qquad   \forall i \in S, \forall j \in A_q 
\label{eq:powerrxalld}
\end{equation}

$P^{t}_{ij}$ and $P^{r}_{ij}$ are equivalent to the expressions seen in Eq. (\ref{eq:powertxall}) and (\ref{eq:powerrxall}). 

Note that now the power dissipation is defined per application $j$ at each node $i$ instead of only per node $i$. This 
differentiation is needed because the remaining activity time of each application in $A_q$ may be different, which will impact on the energy consumption of the node as shown in the following restriction.

Let $X_i$ be a constant equal to 1 if the node $i$ was active before the arrival of the new application, and 0 otherwise. This constant is equivalent to the variable $x_{in-1}$ shown in the general model. Analogously, let $Y_{ijk}$ be a constant equal to 1 if the test point $k$ of the application $j$ was being sensed in the node $i$ just before the arrival of the new application, and 0 otherwise. Thus, this constant is equivalent to the variable $y_{ijkn-1}$ previously shown. $\Delta \tau_j$ is the remaining activity time of application $j$ at time instant $\tau_q$ ($\Delta \tau_j = \tau_j + \epsilon_j - \tau_q$), $E_i(\tau_q)$ is the remaining energy that node $i$ still has at $\tau_q$, and $\lambda_i$ is a variable indicating the residual energy that node $i$ would have once the activity time of the applications deployed in it or forwarded by it expires. Again, sinks do not have energy constraints. With this, the energy constraints that must be ensured in every node are:
\begin{multline}
\sum_{j \in A_q} P^{t}_{ij} \Delta \tau_j +  \sum_{j \in A_q} P^{r}_{ij} \Delta \tau_j +  f\left(\sum_{j \in A_q} \sum_{ k \in W_{j}} y_{ijk} l_j\right) \Delta \tau_j   \\ 
+ \varphi_i \cdot (x_{i}-X_{i})\cdot x_{i} +  \sum_{j \in A_q} \delta_{ij} \sum_{ k \in W_{j}} (y_{ijk}-Y_{ijk})\cdot y_{ijk}   \\ + 
\lambda_i = E_i(\tau_q) \qquad  \forall i \in S'
\label{eq:constf110}
\end{multline}

\noindent with $\lambda_i \geq 0$, which are equivalent to constraints defined in Eq. (\ref{eq:const17}).

\subsection{Objective function}\label{sec:din_obj}

The sets of restrictions described above forces the deployment of the application arriving at time instant $\tau_q$ and also of all the applications that are already active at that moment. If the solution space described by these restrictions is null, then the new application cannot be deployed ensuring the presence of the previous applications and therefore the system rejects it. If the solution space contains several feasible solutions to deploy a new application, we should select a solution that increases the probability of accepting future applications. As stated before, the remaining energy of the nodes is one of the most limiting factors to deploy new arriving applications. Therefore, we propose three possible objective functions for the optimization problem, being all of them related to the residual energy of the network or its nodes. The quality of the three strategies will be analyzed in Section~\ref{performance}.

The first one is to maximize the total residual energy of the network, that is:
\begin{equation}
\max \sum_{i \in S'} \lambda_i
\label{eq:maximiza1} 
\end{equation}

The second one is to maximize the residual energy of the node with the lowest energy,
\begin{equation}
\max \lambda \qquad \lambda \leq \lambda_i \qquad \forall i \in S'
\label{eq:maximiza2} 
\end{equation}

Finally, we also consider a weighted sum of the two previous alternatives:
\begin{equation}
\max \left( \lambda + \frac{1}{|S'|}\sum_{ i \in S'} \lambda_i \right) \qquad \lambda \leq \lambda_i \qquad \forall i \in S'
\label{eq:maximiza3}
\end{equation}

In the following we will denote these three strategies as \textit{Total}, \textit{Max-min} and \textit{Mixed} respectively.

\section{Heuristic algorithm for dynamic resource allocation}
\label{sec:heuristic}
We propose hereafter a heuristic based on a greedy algorithm to solve the dynamic resource allocation problem described in Section \ref{sec:dynamic}. This heuristic is executed each time a new application $a_q$ arrives to the reference area. 

The objective of the heuristic is to deploy the arriving application in a set of sensor nodes that cover all its test points without dropping out any of the applications running at that moment. This implies that the solution found by the heuristic must satisfy the same constraints defined in Section \ref{sec:dynamic}. Additionally, the solution should deploy the new application using the existing resources of the network efficiently, so that the probability of deploying new applications in the future is maximized.

As the most restricting factor to deploy new applications is the spare energy of the nodes, the main idea of the greedy algorithm is to deploy the new application trying to maximize the residual energy of the energy bottleneck node. To do so, the algorithm sorts all the nodes in the set $S_{qk}$ in an decreasing order with respect to a \emph{metric} $e_i$ defined as:
\begin{equation}
e_{i} = \min_{h\in P_{i}} \hat{E}_h(\tau_q)
\label{eq:j_1}
\end{equation}	
\noindent where $P_i$ is the set of nodes forming the path from node $i$ to its corresponding sink (note that these nodes are known in advance since they depend on the routing algorithm used in the network), and $\hat{E}_h(\tau_q$) is the remaining energy that node $h$ would have once all the applications that are deployed on it at time instant $\tau_q$ (when application $a_q$ arrives) leave the network. The term $e_i$ represents therefore the ``bottleneck'' in terms of energy among the nodes in $P_i$. With this, the aim is to deploy the application in the sensor node with the highest value of $e_i$ that has enough free resources to allocate it. 

The detailed pseudocode of the proposed solution is reported in Algorithm 1. The inputs of the heuristic are the available memory and processing capacity of the nodes at time instant $\tau_q$, which are denoted as $M_i(\tau_q)$ and $L_i(\tau_q)$ respectively; the terms $\hat{E}_i(\tau_q)$ defined previously; and the available transmission resources of the links between each node $i$ and its predecessor in the routing graph, which are denoted as $B_i(\tau_q)$. Note that, before any application has arrived to the network, $M_i(\tau_q), L_i(\tau_q)$ and $\hat{E}_i(\tau_q)$ are equal to $M_i, P_i$ and $E_i$, and $B_i(\tau_q)$ are equal to 1 indicating that the whole airtime of all the links are fully available. 

Once the new application $a_q$ has arrived, the first step is to define the variables $M_i^{(r)}, P_i^{(r)}, E_i^{(r)}$ and $B_i^{(r)}$, which are used to store a copy of $M_i(\tau_q), L_i(\tau_q)$, $\hat{E}_i(\tau_q)$ and $B_i(\tau_q)$. These variables allow obtaining the remaining resources of the network if application $a_q$ can be effectively deployed (i.e. if there are enough resources in the network). We also define the lists $S_{qk}'$ to store the nodes that cover each test point $w_k$ of application $a_q$ and have enough resources to sense effectively this test point. Each list $S_{qk}'$ is initialized with the elements in the corresponding set $S_{qk}$ (line 2).

Then, the algorithm enters into its main loop and tries to sense all the test points of $a_q$ following the energy criteria defined previously (lines 3 - 30). To that end, we compute the terms $e_i$ for all the nodes in $S_{qk}'$ using Eq. (\ref{eq:j_1}) and sort the nodes in a decreasing order of $e_i$. Then, we look for the sensor node with the highest value of $e_i$ (sensor node $s_f$ in line 7) and try to deploy $a_q$ on it. For simplicity, this part of the heuristic is detailed in Algorithm 2 and will be explained later. If this is feasible, we update the variables $M_i^{(r)}, P_i^{(r)}, E_i^{(r)}$ and $B_i^{(r)}$ (line 11) and we move to the next test point of $a_q$. If not, the next step depends on whether sensor node $s_f$ is already active or not. If it is not active, $s_f$ is removed from $S_{qk}'$ and the next node with the highest value of $e_i$ is selected (line 27). If it is active, we try to free some of its resources by moving one of the applications it hosts to another node (lines 14-26). 

In order to decide which application to move, we compute for each application $a_j$ on $s_f$ and each node $s_i \in S_{jk} - \{s_f\}$, the cost of moving $a_j$ to $s_i$. This cost is defined as $c(a_j, s_i) = h(s_i) - h(s_f)$, with $h(s_i)$ the number of hops from sensor node $s_i$ to its corresponding sink. We store each pair of application $a_j$ and alternative node $s_i$ on the list $L_f$ and sort it on increasing order of $c(a_j, s_i)$ (lines 14-15). Then, we try to move the first application in this list to its corresponding alternative node ($a_c$ and $s_d$ in line 17) and deploy $a_q$ in $s_f$. For simplicity, this part of the heuristic is detailed in Algorithm 3 and will be explained later. If they fit, we update the variables $M_i^{(r)}, P_i^{(r)}, E_i^{(r)}$ and $B_i^{(r)}$ (line 21) and we move to the next test point of $a_q$. If not, we remove the pair $(a_c, s_d)$ from $L_f$ and repeat the process iteratively until the list is empty. When this happens, we remove $s_f$ from $S_{qk}'$ and the next node with the highest value of $e_i$ is selected.

If after all this process we cannot sense a test point in $W_q$ (i.e. the list $S_{qk}'$ is empty for some $w_k \in W_q$), then the application is rejected and the system remains unchanged in a way that the previous applications can be served. On the contrary, if all the test points are sensed, we deploy the incoming application in the nodes selected during the execution of the algorithm, updating the resources of the involved nodes (i. e., the terms $M_i(\tau_q)$, $L_i(\tau_q)$, $\hat{E}_i(\tau_q)$ and $B_i(\tau_q)$).

\begin{algorithm}[!t]
	
	\caption{Pseudo-code of the dynamic greedy solution}
	\label{alg:dynamic_algorithm}
	\begin{algorithmic}[1]
		
		\State Application $a_q$ arrives at time $\tau_q$
		\State Initialize variables $M_i^{(r)}, P_i^{(r)}, E_i^{(r)}, B_i^{(r)}$ and lists $S_{qk}'$
		
		\ForAll {$w_k \in W_q$}
		\State Compute $e_i$ with (\ref{eq:j_1}) $\forall s_i \in S_{qk}'$
		\State Sort all the nodes in $S_{qk}'$ in decreasing order of $e_i$
		
		\Repeat
		
		\State $s_f \gets \arg \max_{s_i \in S_{qk}'} (e_i)$		
		\State Check if $a_q$ fits into $s_f$ (alg. 2)
		
		\If {it fits}
		\State Test point $w_k$ is sensed
		\State Update variables $M_i^{(r)}, P_i^{(r)}, E_i^{(r)}, B_i^{(r)}$
		\Else 
		\If {$s_f$ is active}
		\State Store in $L_f$ all the pairs $(a_j, s_i)$, with $a_j$ \phantom{abcabcabcabc} on $s_f$ and $s_i \in S_{jk} - \{s_f\}$ 
		\State Sort the elements of $L_f$ in increasing order \phantom{abcabcabcabc} of $c(a_j, s_i)$
		
		\Repeat 
		\State $(a_c, s_d) \! \gets \! \arg \max_{(a_j, s_i) \in L_f} (c(a_j, s_i)) $
		\State Check if $a_c$ fits into $s_d$ and $a_q$ fits into \phantom{abcabcabcabcabc.} $s_f$ without $a_c$ (alg. 3)
		\If {they fit}
		\State Test point $w_k$ is sensed
		\State Update vars. $M_i^{(r)}, P_i^{(r)}, E_i^{(r)}, B_i^{(r)}$
		\Else
		\State Remove $(a_c, s_c)$ from $L_f$
		\EndIf
		\Until {$w_k$ is sensed or $L_f$ is empty}
		\EndIf
		\State Remove $s_f$ from $S_{qk}'$
		\EndIf	
		\Until {$w_k$ is sensed or $S_{qk}'$ is empty}
		\EndFor				
		\If {all $w_k \in W_q$ are sensed}
		\State Deploy $a_q$ and update $M_i(\tau_q)$, $L_i(\tau_q)$, $\hat{E}_i(\tau_q)$, $B_i(\tau_q)$
		\Else
		\State Reject $a_q$
		\EndIf
		
	\end{algorithmic}
	
\end{algorithm}

Now, we describe the procedure followed to determine if the application $a_q$ fits into the sensor node $s_f$ (Algorithm 2). First, we define the variables $M_i^{(a)}, P_i^{(a)}, E_i^{(a)}$ and $B_i^{(a)}$, which are used to store a copy of $M_i^{(r)}, P_i^{(r)}, E_i^{(r)}$ and $B_i^{(r)}$ and are initialized with their values. These variables allow obtaining the remaining resources of the nodes if $a_q$ can be effectively deployed on $s_f$.

To check this, we have to compute the remaining resources of all the nodes that would be affected if the application were deployed on $s_f$. Namely: 
\begin{itemize}
	\item The remaining available memory and processing capacity of node $s_f$ (lines 2-3)
	\item The remaining energy of $s_f$. To compute it, we have to remove the energy that $s_f$ requires to sense the test point, which corresponds to the term $\epsilon_q f(l_q)$, and the energy required to send the gathered data to the next node in the path to its sink, which corresponds to the term $\epsilon_q c_q (\beta_1 + \beta_2 d_{f\rightarrow n(f)}^\gamma) $. In this expression $n(f)$ is the node corresponding to the next hop of $s_f$ in the path calculated by the routing algorithm, and $d_{f\rightarrow n(f)}$ is the distance between $s_f$ and $s_{n(f)}$ (line 6). Additionally, if $s_f$ is off, we also have to consider the energy required to power it on (line 11).
	
	\item The remaining energies of the rest of nodes in the path $P_f$. To compute them, we have to remove for each node $s_g$ in the path (different from $s_f$) the energy that it requires to receive the sensed data from the previous node in the path, which corresponds to $\epsilon_q  c_q \rho$, and the energy required to retransmit them to the next node, which corresponds to $\epsilon_q c_q \left( \beta_1 + \beta_2 d_{g\rightarrow n(g)}^\gamma \right)$ (line 8). Again, if $s_g$ is off, we also have to consider the energy required to power it on (line 11).
	
	\item The remaining transmission resources of each link of the path $P_f$ (line 13). We name as $l_g$ (or directly $g$) the link between the node $s_g$ and the node being its next hop $s_{n(g)}$. For each link, the required transmission resources correspond to the airtime needed to transmit the sensed data, which is $c_q/C_g$.
	
	\item The remaining transmission resources of the set of links that interfere or are interfered by link $l_g$ (lines 14-15). We name this set as $I_g$ and it is formed by the links whose receiver is within the interference range of the node $s_g$ and the links where the node $s_{n(g)}$ is within the interference range of its transmitter. % Every time the link $l_g$ is transmitting, the links in $I_g$ must remain silent so that there is no interference between them (and conversely, every time a link in $I_g$ is transmitting, $l_g$ must be silent). Therefore, the transmission time of link $l_g$ is shared among all the links in $I_g$.
	
\end{itemize}

If all the auxiliary variables $M_i^{(a)}, P_i^{(a)}, E_i^{(a)}$ and $B_i^{(a)}$ are higher than zero, we can state that $a_q$ fits into $s_f$ (line 21).

\begin{algorithm}[!t]
	\caption{Check if application $a_q$ fits into the node $s_f$}
	\label{alg:fits1}
	\begin{algorithmic}[1]
		
		\State Initialize variables $M_i^{(a)}, P_i^{(a)}, E_i^{(a)}, B_i^{(a)}$

		\State $M_f^{(a)} \gets  M_f^{(a)} - m_q$
		\State $P_f^{(a)} \gets  P_f^{(a)} - p_q$
		
		\ForAll {$s_g \in P_f$}

		\If {$s_g = s_f$}
		\State $E_f^{(a)} \! \gets \! E_f^{(a)} - \epsilon_q \left( f(l_q) + c_q \left(\beta_1 + \beta_2 d_{f\rightarrow s(f)}^\gamma\right) \! \right)$
		\Else		
		\State $E_g^{(a)} \gets E_g^{(a)} - \epsilon_q c_q \left( \rho + \beta_1 + \beta_2 d_{g\rightarrow s(g)}^\gamma \right) $
		\EndIf
		
		\If {$s_g$ is off}
		\State $E_g^{(a)} \gets E_g^{(a)} - \varphi_g$
		\EndIf
		
		\State $B_g^{(a)} \gets B_g^{(a)} - c_q / C_g$
		
		\ForAll {$l_h \in I_g$}
		\State $B_h^{(a)} \gets B_h^{(a)} - c_q / C_g$
		\EndFor
		\EndFor
		
		\If {any variable $M_i^{(a)}, P_i^{(a)}, E_i^{(a)}, B_i^{(a)}$ is $< 0$}
		\State	\Return $a_q$ does not fit into $s_f$
		\EndIf
		
		\State \Return $a_q$ fits into $s_f$ and $M_i^{(a)}, P_i^{(a)}, E_i^{(a)}, B_i^{(a)}$
		
	\end{algorithmic}
	
\end{algorithm}

Finally, we describe the procedure followed to determine if it is possible to deploy application $a_q$ into the sensor node $s_f$ when we move application $a_c$ from $s_f$ to $s_d$ (Algorithm 3). This procedure consists of two parts, being each of them very similar to the procedure described in Algorithm 2. In the first one, we check if sensor node $s_d$ has enough resources to host application $a_c$ (lines 4-17). The only difference with respect to the verifications done in Algorithm 2 is that the time that the application is in the network is not $\epsilon_c$ but $\tau_c + \epsilon_c - \tau_q$, which impacts on the energy requirements of the nodes (lines 6-8), as well as the energy impact of sending the bytecode of $a_c$ to $s_d$. In the second one, we check if sensor node $s_f$ can host $a_q$ once the resources used by $a_c$ are freed from $s_f$ (lines 20-33). In this case, the difference with respect to Algorithm 2 is that we must add the new available resources that were previously used by $a_c$. 

\begin{algorithm}[!t]
	\caption{Check if $a_q$ fits into the node $s_f$ when $a_c$ is moved to $s_f$}
	\label{alg:fits2}
	\begin{algorithmic}[1]
		
		\State Initialize variables $M_i^{(a)}, P_i^{(a)}, E_i^{(a)}, B_i^{(a)}$

		\State $M_d^{(a)} \gets  M_d^{(a)} - m_c$
		\State $P_d^{(a)} \gets  P_d^{(a)} - p_c$
		
		\ForAll {$s_g \in P_d$}

		\If {$s_g = s_d$}
		\State $E_d^{(a)} \gets E_d^{(a)} - \delta_{cd} $ \phantom{aaaaaaaaaaaaaaaaaaaaaaaaaaaa} \phantom{aaaaaaaa} $- \left( \tau_c + \epsilon_c - \tau_q \right) \!  \left( f(l_c) +  c_c \! \left(\beta_1 + \beta_2 d_{d\rightarrow s(d)}^\gamma\right) \! \right) $
		\Else		
		\State $E_g^{(a)} \gets  E_g^{(a)}$ \phantom{aaaaaaaaaaaaaaaaaaaaaaaaaaaaaaaaa} \phantom{aaaaaaaaaaaaa} $ - c_q \left( \tau_c + \epsilon_c - \tau_q \right) \left( \rho + \beta_1 + \beta_2 d_{g\rightarrow s(g)}^\gamma \right) $
		\EndIf
		
		\If {$s_g$ is off}
		\State $E_g^{(a)} \gets E_g^{(a)} - \varphi_g$
		\EndIf
		
		\State $B_g^{(a)} \gets B_g^{(a)} - c_c / C_g$
		\ForAll {$l_h \in I_g$}
		\State $B_h^{(a)} \gets B_h^{(a)} - c_c / C_g$
		\EndFor
		\EndFor

		\State $M_f^{(a)} \gets  M_f^{(a)} - m_q + m_c$
		\State $P_f^{(a)} \gets  P_f^{(a)} - p_q + p_c$
		
		\ForAll {$s_g \in P_f$}
		
		\If {$s_g = s_f$}
		\State $E_f^{(a)} \! \gets \! E_f^{(a)}   -   \epsilon_q \!  \left( \!  f(l_q)  +   c_q \! \left(\beta_1 + \beta_2 d_{f\rightarrow s(f)}^\gamma\right) \!  \right)$ \phantom{aaaaaaa.} $ +  \left( \tau_c + \epsilon_c - \tau_q \right) \! \left( f(l_c) + c_c \left(\beta_1 + \beta_2 d_{f\rightarrow s(f)}^\gamma\right) \! \right) $
		\Else		
		\State $E_g^{(a)} \gets E_g^{(a)} - \epsilon_q  c_q \left( \rho + \beta_1 + \beta_2 d_{g\rightarrow s(g)}^\gamma \right) $ \phantom{aaaaaa} \phantom{aaaaaaaaaaaaaa} $ +   c_c \left( \tau_c + \epsilon_c - \tau_q \right) \! \left( \rho + \beta_1 + \beta_2 d_{g\rightarrow s(g)}^\gamma \right)$
		\EndIf	
		
		\State $B_g^{(a)} \gets B_g^{(a)} - (c_q - c_c) / C_g $
		
		\ForAll {$l_h \in I_g$}
		\State $B_h^{(a)} \gets B_h^{(a)} - (c_q -c_c) / C_g$
		\EndFor
		\EndFor
		
		\If {any variable $M_i^{(a)}, P_i^{(a)}, E_i^{(a)}, B_i^{(a)}$ is $< 0$}
		\State \Return $a_q$ does not fit into $s_f$
		\EndIf
		
		\State \Return $a_q$ fits into $s_f$ and $M_i^{(a)}, P_i^{(a)}, E_i^{(a)}, B_i^{(a)}$
		
	\end{algorithmic}
	
\end{algorithm}

\section{Performance Evaluation} \label{performance}

In this section, we evaluate in detail the performance of the proposed strategies. Unless otherwise stated, results have been obtained by solving the optimization models of Sections \ref{sec:optim} and \ref{sec:dynamic} using CPLEX software \cite{cplex}. The simulated scenarios consider several area sizes and different number of applications and nodes. In all the cases, results have been obtained averaging the outcome of 100 realizations. We consider a default sensing range of $R^{s}_{i}$ = 40 m for all the sensors \cite{Chen:2007} and a two-ray ground path loss model with $\gamma = 4$ and $g_0 = 8.1 \cdot 10^{-3}$ \cite{Suh07}. $P_{max}$ is set to $-10$ dBm and the receiver sensitivity  $\alpha$ is fixed to $-92$ dBm \cite{cc2420}, which implies a maximum transmission range $R^T_{max}$ of 33 m. Similarly, the interference sensitivity $\mu$ is set to $-104$ dBm, which implies a maximum interference range $R^I_{max}$ of 67 m. 

\subsection{Applications and Sensor nodes}
\label{apsssensornodes}

As a reference, we have focused on \emph{multimedia} applications, which require the sensing, processing and delivery of multimedia content (images and video). Specifically, we consider visual sensor networks, i.e. WSNs designed to perform visual analysis (e.g. object recognition) \cite{RedondiTMC2016}. In that work, a detailed characterization of transmission rates and energy consumption for these applications is provided. Based on this analysis, the transmission rate is set to 12 kb/s, the required memory to 842~Kbytes, the processing load to 69.23 MIPS and the associated power dissipation (function $f$ in eq. (\ref{eq:const17a}) and eq.~(\ref{eq:constf110})) to 0.2 W. Details on how these numbers have been derived are shown in \cite{Delgado2016}. Thus, the requirement vector for visual applications is $o_j = \left\{\text{12 kb/s, 842 KB, 69.23 MIPS}\right\}$. 

To support these visual applications, we consider \emph{high-level} sensor node hardware. Using as a reference BeagleBone platforms \cite{beagle} or similar, the sensor nodes are assumed to have a 720 MHz super-scalar ARM Cortex-A8 processor (up to 720 MIPS) and 256 MB of RAM. In addition they are equipped with an IEEE 802.15.4 radio with an integrated antenna and a low-power USB camera. The reference resource vector is  $O_i = \left\{\text{250 kb/s, 256 MB, 720 MIPS, 32400 J}\right\}$. The energy budget for all the nodes but sinks, which we assume that can be plugged directly into the grid, is 32400 J assuming that a node runs at 3 V with 3 Ah of battery supply (2 AA batteries).

\subsection{General optimization framework results}

We begin by assessing the general optimization framework of Sec.~\ref{sec:optim}. The solution of this problem gives an upper-bound to the performance of the strategies evaluated in the following subsection and gives a hint on the value of ``future information'' (i.e. the knowledge about the applications that have not arrived to the system yet). Nevertheless, this model is neither realistic nor easy to solve computationally due to the extremely high number of variables needed to solve it.

Since the number of variables needed to solve the general problem increases quickly, we need to use a smaller scenario than the used in the following subsection. We have considered a scenario formed by 18 BeagleBone nodes, where 100 visual applications are generated according to a Poisson process with rate of $0.5$ applications per hour and a constant activity time $\epsilon_j$ of $5$ hours. We assume that each sensor is able to cover $N_{ij} = 1$ test point of the same application and that there is $1$ sink node in the network. Sensor nodes are deployed in a $100 \times 100$ m scenario.  

%Table \ref{table:applications_strategiesA} gives the number of deployed applications achieved for the previous strategies (\emph{Only Restrictions (O.R.), Total, Max-min, Mixed and Heuristic}) and the general optimization strategy seen in Section~\ref{sec:optim} (\emph{Global}). 

Table \ref{table:applications_strategiesA} gives the number of deployed applications achieved for the general optimization strategy seen in Section~\ref{sec:optim} (\emph{Global}), the different proposed strategies with the dynamic model (\textit{Total}, \textit{Max-min} and \textit{Mixed}) and the heuristic algorithm (\textit{Heuristic} in the legends). In addition, results are also presented for the case where no objective function is considered, and the dynamic optimization problem only tries to deploy the application satisfying all the restrictions (\textit{Only restrictions} (O.R.) in the legends).

The number of test points for each application is $1$ or $2$. As can be seen, the improvement of solving the \emph{Global} problem is not significant, so the proposed solutions for optimizing online the energy of the network work properly.

%\begin{center}
\begin{table} [!t]
\renewcommand{\arraystretch}{1.3}
\centering
\caption{Deployed applications for each strategy}
\begin{footnotesize}
	\centering
	\begin{tabular}{| r || c | c | c | c | c | c |}
 		\hline
 		\multirow{2}{*}{\textbf{}} & \multicolumn{6}{c|}{\textbf{Strategy}} \\ \cline{2-7}
& O.R. & Total & Max-min & Mixed & Heuristic & Global \\ \hline \hline
			%1 Test Point per Application & 97.85 & 99.65 & 99.65 & 99.65 & 99.64 & 99.94 \\ 
			1 TP. A.& 97.85 & 99.65 & 99.65 & 99.65 & 99.64 & 99.94 \\ 
\hline
%2 Test Points per Application & 76.71 & 75.36 & 91.71 & 95.64 & 87.64 & 98.57 \\ 
2 TP. A.& 76.71 & 75.36 & 91.71 & 95.64 & 87.64 & 98.57 \\ 
\hline
\multicolumn{7}{l}{{\scriptsize \emph{TP. A. Test Point per Application}}}
    	\end{tabular}
\end{footnotesize}
\label{table:applications_strategiesA}
\end{table}
%\end{center} 

\subsection{Dynamic resource allocation results}

The following results have been obtained by solving the optimization model of Sec.~\ref{sec:dynamic} and the heuristic algorithm proposed in Sec.~\ref{sec:heuristic}. We have considered a scenario formed by 36 BeagleBone nodes as a reference example to thoroughly evaluate the validity of the dynamic model, the performance of the different objective functions proposed in Sec.~\ref{sec:din_obj} and the heuristic greedy algorithm. In each realization, 200 visual applications are generated according to a Poisson process with rate of $1$ application per hour and a constant activity time $\epsilon_j$ of $5$~hours. The number of test points is $3$ for each application, and we assume that each sensor is able to cover $N_{ij}=1$ test point of the same application and that there are $2$ sink nodes in the network. Sensor nodes are deployed in a $141 \times 141$~m scenario.  

\begin{figure*}[!t]
	\centering{
		\subfloat[]{\includegraphics[width=2.3in]{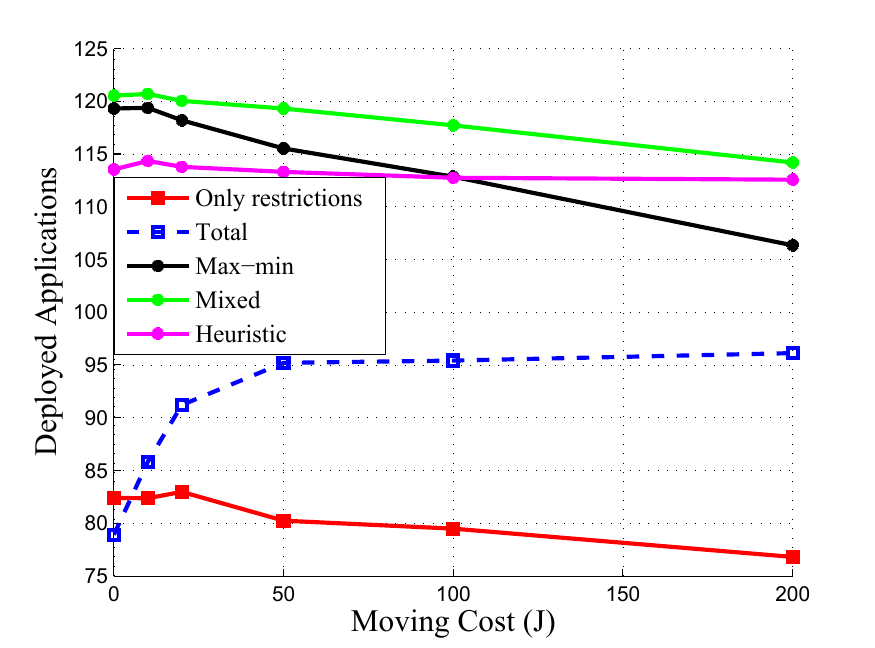}	\label{fig:ACONApps}} \hfil
		\subfloat[]{\includegraphics[width=2.3in]{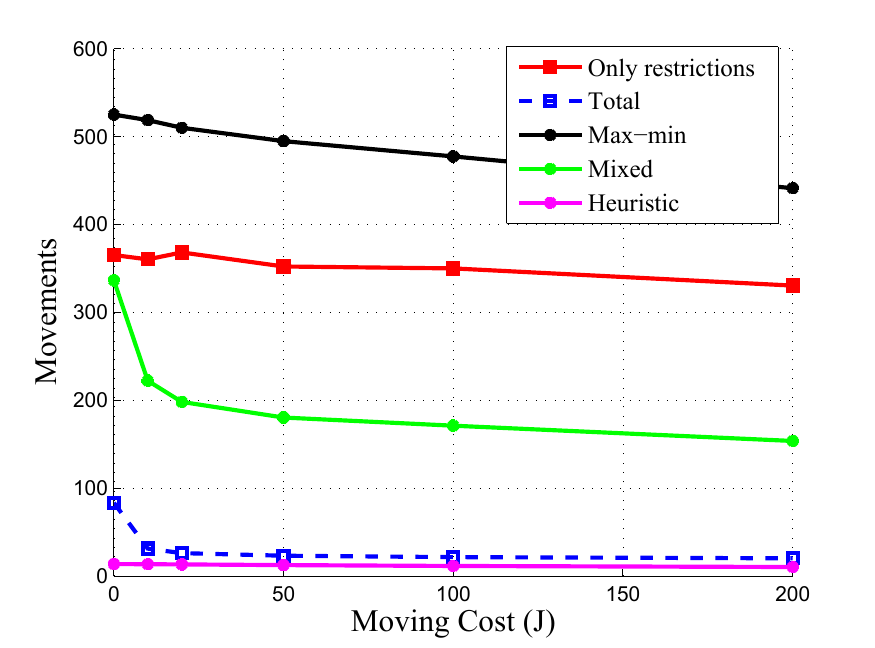}	\label{fig:ACMov}} \hfil
		\subfloat[]{\includegraphics[width=2.3in]{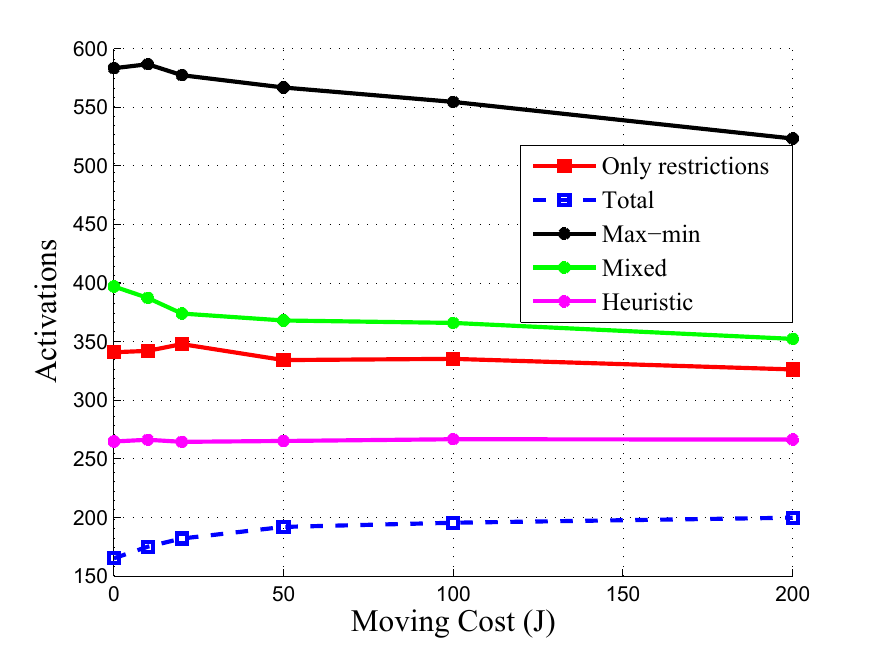} \label{fig:ACActiv}}
		
		\caption{Impact of moving cost. a) Deployed applications. b) Number of movements. c) Number of activations. $\varphi_i$ = 10~J $\forall i \in S$.}	\label{fig:AC}
	}	
\end{figure*}
	
\begin{figure*}[!t]
		\centering
		{
			\subfloat[]{\includegraphics[width=2.3in]{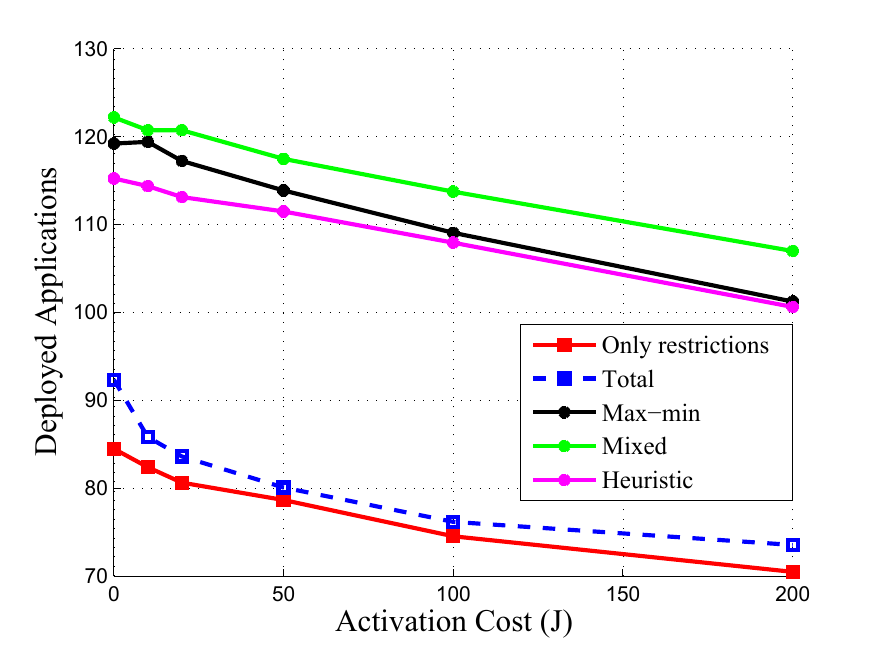}
				\label{fig:SCONApps}}
			\subfloat[]{\includegraphics[width=2.3in]{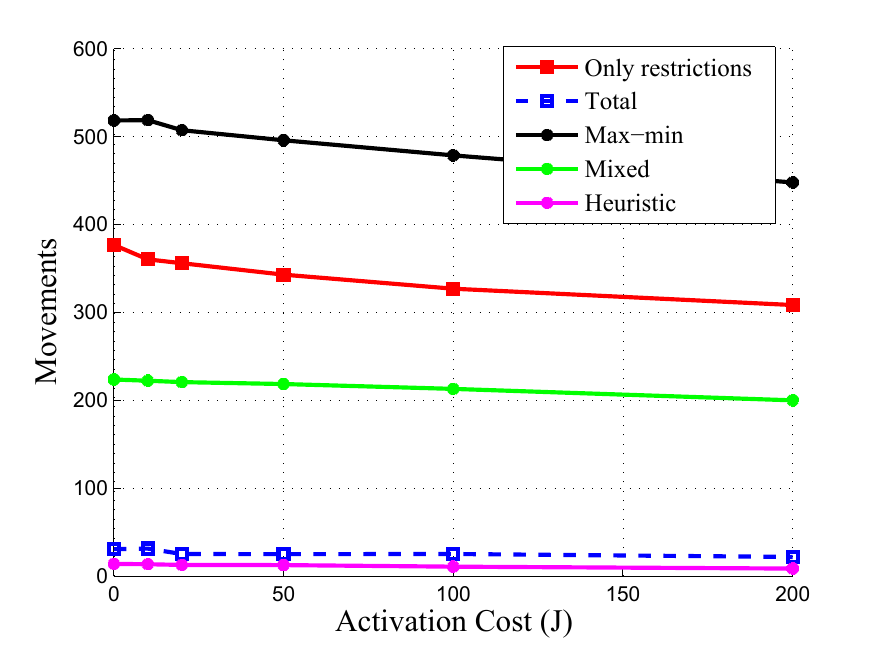}
				\label{fig:SCMov}}
			\subfloat[]{\includegraphics[width=2.3in]{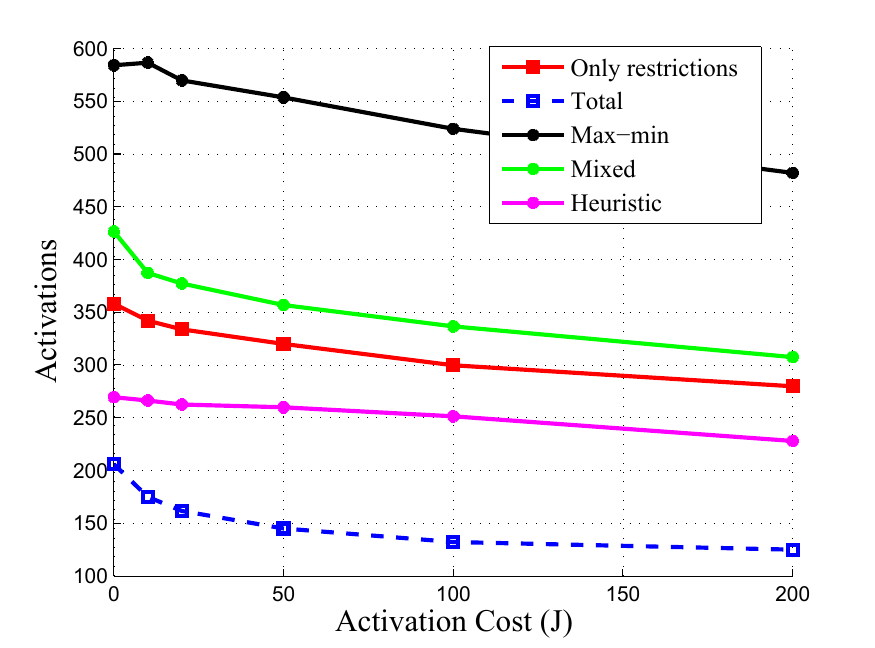}
				\label{fig:SCActiv}}
			
			\caption{Impact of activation cost. a) Deployed applications. b) Number of movements c) Number of activations. $\delta_{ij}$ = 10~J $\forall i \in S, \forall j \in A$.}
			\label{fig:SC}
			
		}
	\end{figure*}

\begin{figure}[!t]
	\centering
	{
		\subfloat[]{\includegraphics[width=2.3in]{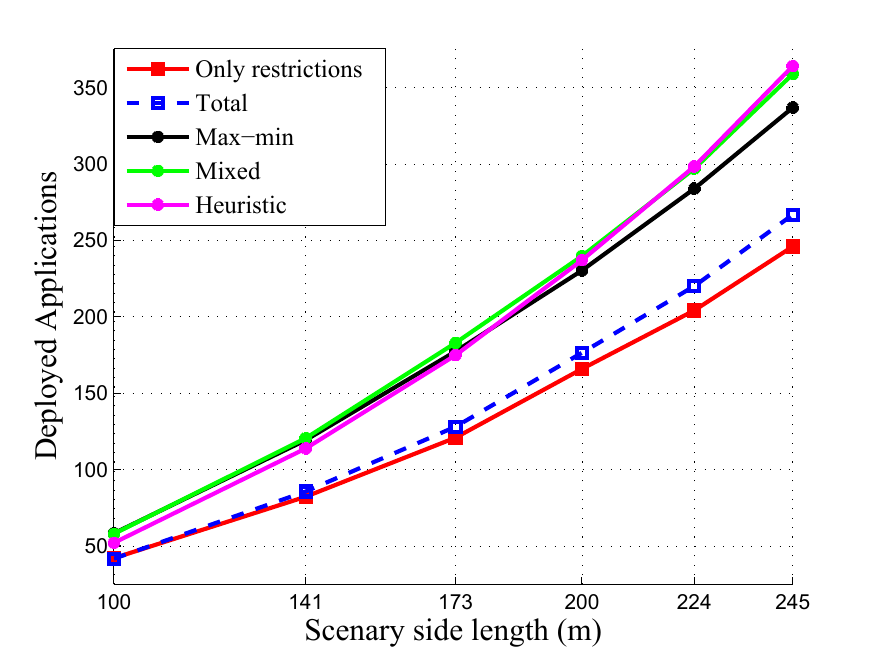}
			\label{fig:appheu}}\\
		\subfloat[]{\includegraphics[width=2.3in]{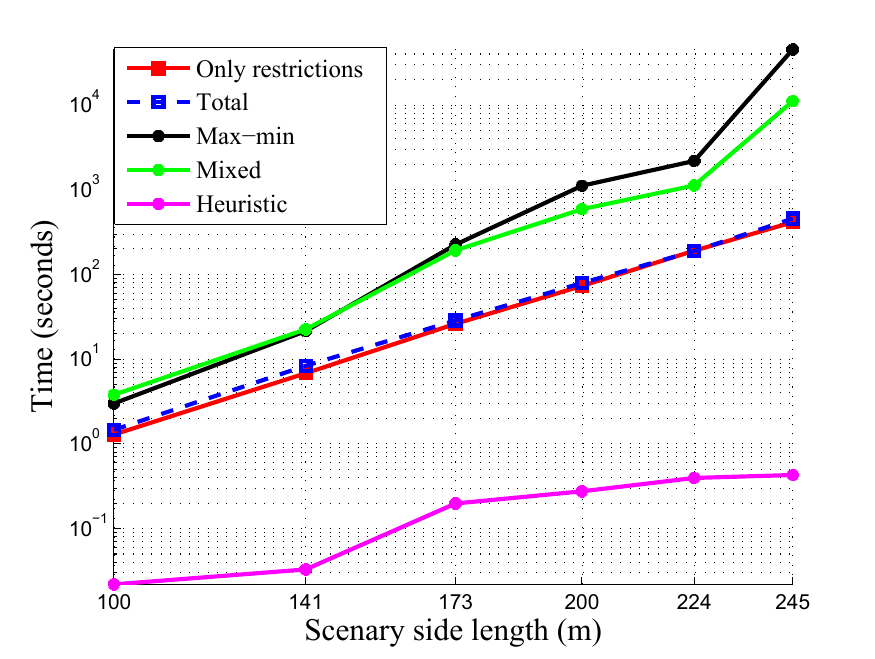}
			\label{fig:timeheu}}
		
		\caption{Performance evaluation of the heuristic algorithm vs the dynamic scheme. (a) Deployed applications (b) Computation Time (in log scale)}
		\label{fig:apptimeheu}
	}
\end{figure}

\begin{figure*}[!b]
	\centering
	{
	\subfloat[]{\includegraphics[width=2.3in]{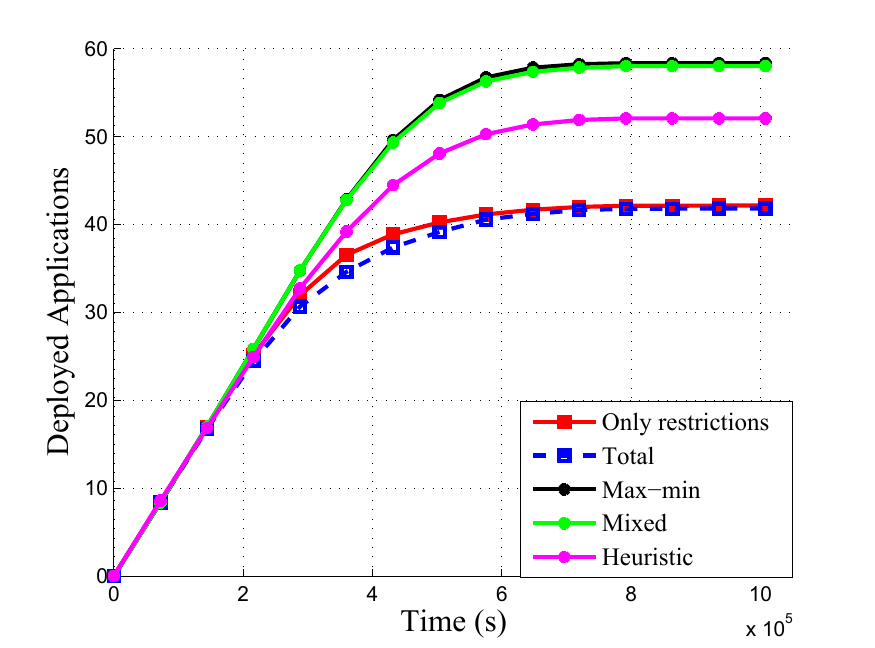}\label{fig:app_100x100}}\hfil
	\subfloat[]{\includegraphics[width=2.3in]{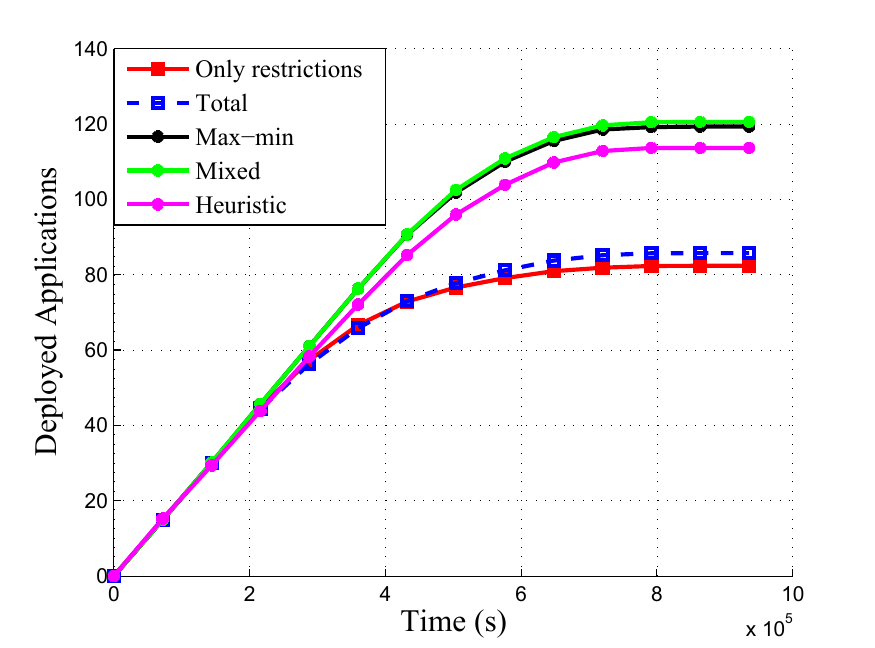}	\label{fig:app_141x141}	}\hfil
	\subfloat[]{\includegraphics[width=2.3in]{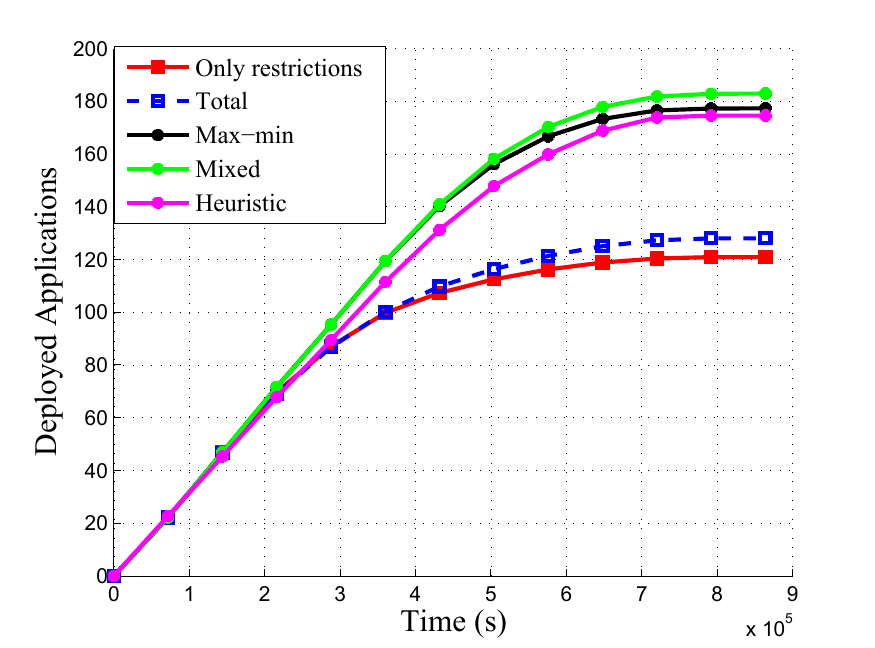} \label{fig:app_173x173}}\\
	\subfloat[]{\includegraphics[width=2.3in]{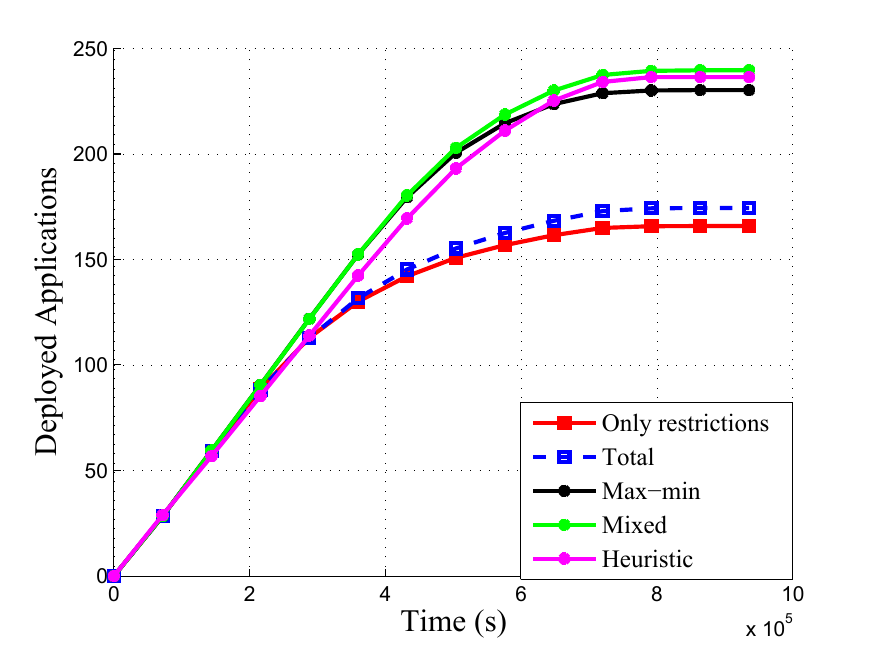}\label{fig:app_200x200}}	\hfil
	\subfloat[]{\includegraphics[width=2.3in]{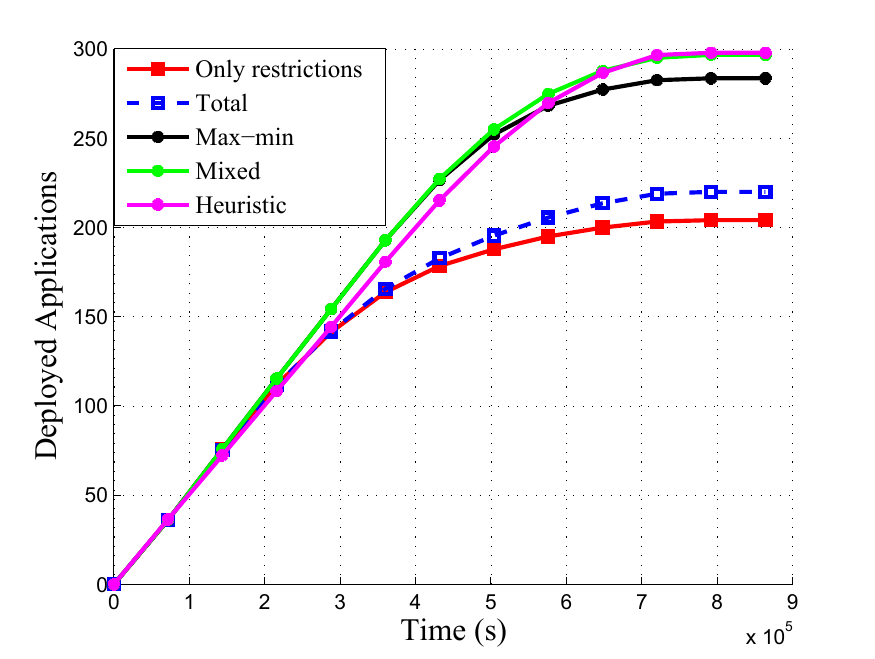}\label{fig:app_224x224}}\hfil
	\subfloat[]{\includegraphics[width=2.3in]{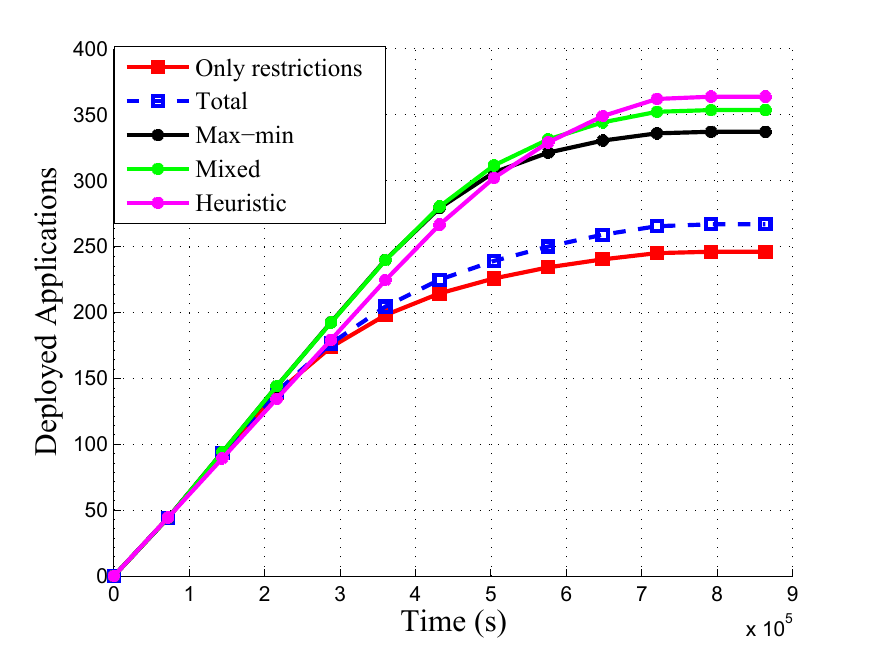}\label{fig:app_245x245}}
	
	\caption{Temporal evolution of the deployed applications for different scenarios. (a) 100 x 100 m (b) 141 x 141 m (c) 173 x 173 m (d) 200 x 200 m (e)  224 x 224 m (f) 245 x 245 m}
	\label{fig:ev_temporal}
}
\end{figure*}

Figs.~\ref{fig:AC} and \ref{fig:SC} show the impact of the energy cost of moving an application $j$ from one node $h$ to another one $i$, $\delta_{ij}$, and the energy cost of activating a node $i$, $\varphi_i$, for the different proposed strategies (\textit{Total}, \textit{Max-min}, \textit{Mixed}, \textit{Heuristic} and \textit{Only restrictions}). Fig.~\ref{fig:ACONApps} shows that the total number of actually deployed applications in the system is higher for the \textit{Mixed} approach for all the moving cost values; as expected the worst results are obtained with the \textit{Only restrictions} approach. The \textit{Max-min} and the greedy heuristic algorithm provide results close to the \textit{Mixed} approach. The near-flat behaviour of the heuristic can be explained by the extremely low number of movements it performs: as there are barely any movement, the impact of the moving cost is negligible. More details about this behavior are given later, when explaining Fig.~\ref{fig:apptimeheu} and Fig.~\ref{fig:ev_temporal}.  On the other hand, when the total energy is maximized, the number of deployed applications rises with higher moving costs. This can be explained as follows: when the moving cost is low and a new application arrives at the system, it is preferred to move one current application from one active node to another, rather than activating a new node so as to maximize the overall residual energy. This makes new applications tend to be located in the activated nodes, making the energy of these nodes be spent faster. In the end, this leads to nodes running out of energy earlier, making the network disjoint and reducing the number of applications that can be deployed. 

Figs.~\ref{fig:ACMov} and \ref{fig:ACActiv} show that \textit{Only restrictions} and \textit{Max-min} are the strategies that have more movements and activations. This is straightforward for \textit{Only restrictions}, since applications are deployed without any additional objective rather than fulfilling the constraints, and therefore the specific nodes where the applications are deployed are chosen randomly as long as the constraints are satisfied. A similar explanation can be applied to the \textit{Max-min} strategy: equations (\ref{eq:constf110}) and (\ref{eq:maximiza2}) only consider the node with the lowest energy, so the remaining nodes can be activated or receive an application without any penalty in the objective function. On the other hand, the lowest number of movements is obtained with the heuristic algorithm since in this case the number of possible movements that are tested is the lowest: we only move an application from one node to another so as to put the arriving application in its instead.

Finally, it must be noted (Fig.~\ref{fig:ACActiv}) that the number of activations remains almost constant (and not rises) when the moving cost increases for the \textit{Only restrictions}, \textit{Mixed} and \textit{Heuristic} strategies. This is because the activation of a new node also implies that this node has to receive the bytecode of the application, so the higher moving cost cannot be compensated by activating more nodes. However, for the \textit{Total} strategy, the number of activations increases as the moving cost increases. As noted in the explanation of Fig.~\ref{fig:ACONApps}, for low moving costs, this solution tends to move applications between active nodes instead of activating new nodes.

Fig.~\ref{fig:SCONApps} shows that the \textit{Mixed} strategy keeps providing the best performance in terms of deployed applications for different values of the activation cost. Again, for the same reasons explained above, \textit{Max-min} is the strategy with more movements and activations (Figs~\ref{fig:SCMov} and \ref{fig:SCActiv}). In addition, it is worth noting again that for \textit{Heuristic}, the number of movements and activations keeps almost constant as the activation cost increases.

\begin{figure}[!t]

%\vspace{-0.2cm}
	\centering
	{
		\subfloat[]{\includegraphics[width=2.3in]{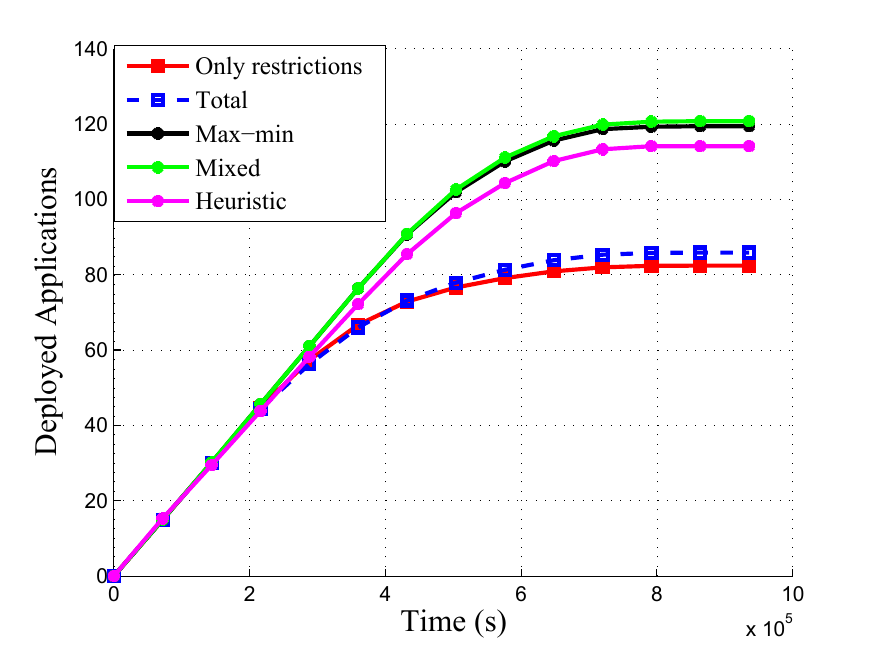}
			\label{fig:1010apps}}\\
		\subfloat[]{\includegraphics[width=2.3in]{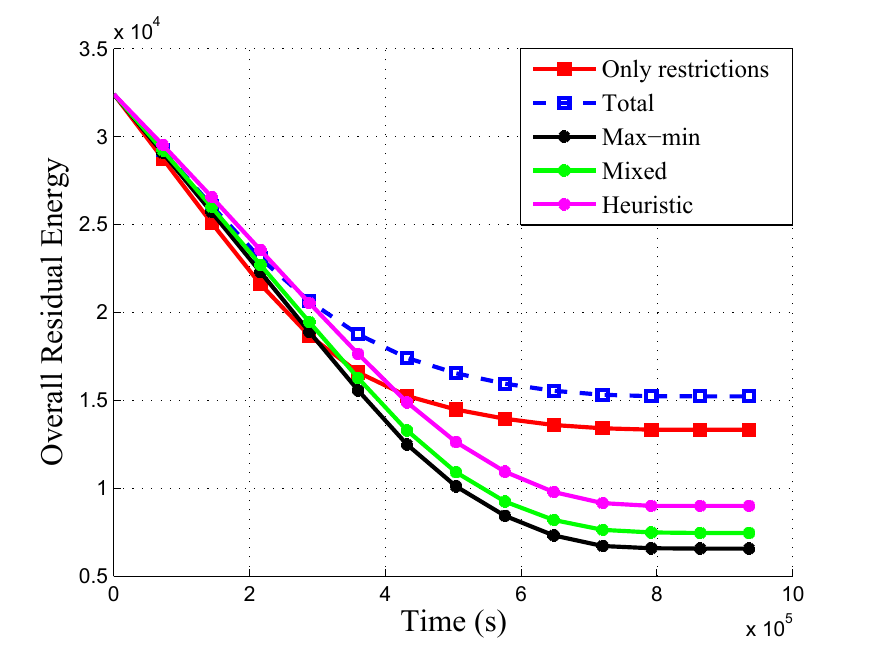}
			\label{fig:1010E}}
		
		\caption{Temporal evolution of system performance. a) Deployed applications. b) Overall residual energy. $\varphi_i$ = 10 J. $\delta_{ij}$ = 10 J. $\forall i \in S, \forall j \in A$ }
		\label{fig:1010}
	}
\end{figure}

\begin{figure}[!b]

%\vspace{-1cm}
	\centering
	{
	\includegraphics[width=2.3in]{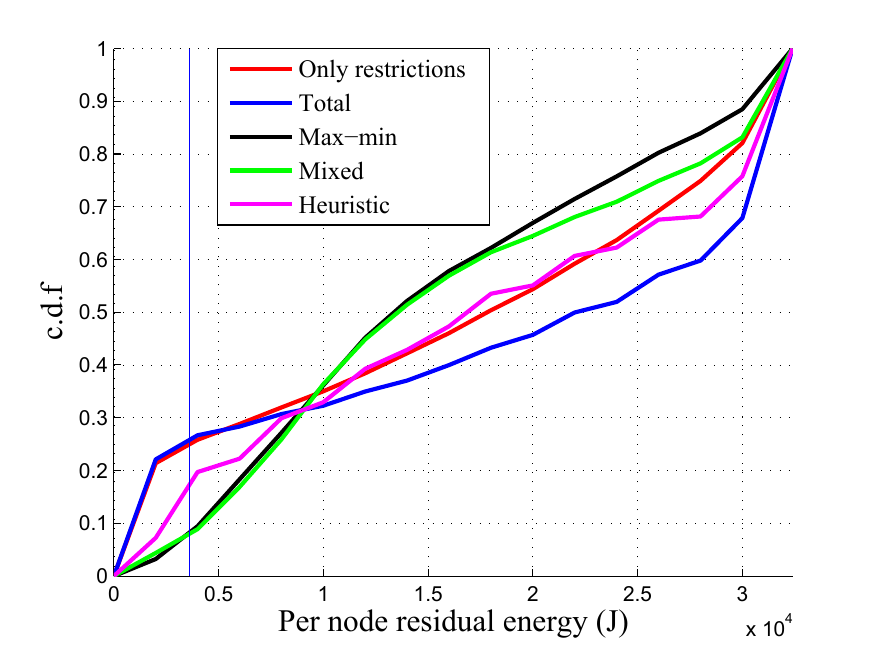}
		
		\caption{Cumulative distribution function of the per node residual energy at time 360000.}
		\label{fig:cdf}
	}
\end{figure}

In Figs.~\ref{fig:1010} and \ref{fig:cdf} we focus on the temporal evolution of the system fixing both $\varphi_{i}$ and $\delta_{ij}$ to 10 J. As can be seen in Fig.~\ref{fig:1010apps}, at first (when all the nodes have a high residual energy), applications are equally deployed for all the strategies. However, as time passes and energy depletes, the fairer energy distribution obtained by the \textit{Mixed} and \textit{Max-min} strategies allows deploying more applications in the system. Comparing Figs.~\ref{fig:1010apps} and \ref{fig:1010E}, an inverse relationship between the residual energy and the deployed applications is observed. Therefore, the key to design a proper strategy is not strictly maximizing the total energy of the network (which is done with the \emph{Total} strategy), but to ensure that the network has enough energy to keep being a connected graph as applications are deployed.

In Fig.~\ref{fig:cdf} the cumulative distribution function of the residual energy per node at time 360000 s is shown. The vertical blue line represents the minimum energy required for a node to sense an application and send the data at the maximum transmission power. As can be seen, the probability of a node not having enough energy to admit a new application is the lowest for th \textit{Mixed} and \textit{Max-min} strategies, which confirms the results shown in Fig.~\ref{fig:1010apps}.
  
In order to assess the performance of the proposed solutions for different scenario sizes, we vary now the size of the reference network topology while maintaining the same node density and a maximum transmitted power of $P_{max}$~=~$-10$~dBm. Both the activation and the moving costs are set to 10~J. Table \ref{table:scenaries} summarizes the main features of the tested scenarios. For example, the first one is a $100\times 100$ m scenario with 18 BeagleBones (one of them acting as a sink) where 100 ATC applications are generated according to a Poisson process with rate of $0.5$ applications per hour and a constant activity time $\epsilon_j$ of $5$ hours. In all the cases, the number of test points per application is $3$, and we assume that each sensor is able to cover $N_{ij} = 1$ test points of the same application.

Fig.~\ref{fig:apptimeheu} depicts the number of deployed applications and the solution time of the proposed solutions. The results have been obtained on 3.0 GHz Quad Core Intel Woodcrest (64~bits) machines with 8 GB RAM and 250 GB SATA storage, averaging over 100 randomly generated network topologies for each scenario of Table \ref{table:scenaries}. These results show that for all the scenarios, the computation time of the \textit{Heuristic} approach is much lower. According to the number of deployed applications, the \textit{Heuristic} approach is always close to the \textit{Mixed} and \textit{Max-min} strategies. Therefore, we can consider that the proposed heuristic achieves near optimal results for all the scenarios considered so far (varying scenario sizes, moving cost and activation size), with a negligible computational cost. 

Additionally, in Fig.~\ref{fig:appheu} we notice that for large scenarios, the deployed applications with the heuristic algorithm get slightly higher values than the other strategies. This can be explained as follows: in some cases, the heuristic algorithm rejects arriving applications that could have been deployed if a more exhaustive search had been performed. Typically, these rejected applications tend to consume more network resources than the accepted ones, so the rejections also save more resources for subsequent applications. Thus, later on, when nodes have little residual energy, new applications consuming few resources are more likely to be deployed with the heuristic algorithm. This effect of rejecting some high resource consuming applications to later accept more low resource consuming ones is more evident as the total number of applications rises, in large scenarios. This behaviour is confirmed in Fig.~\ref{fig:ev_temporal}, where the temporal evolution of the number of deployed applications is shown. For large scenarios (Figs.~\ref{fig:app_224x224} and \ref{fig:app_245x245}), the \textit{Heuristic} goes below the \textit{Mixed} strategy at first (for $t<6 \cdot 10^5$) because the \emph{Heuristic} does not find some feasible solutions that the \emph{Mixed} does since the \emph{Heuristic} does not examine the entire solution space whereas the \emph{Mixed} strategy solves the optimization problem, and therefore, always finds the feasible solution, if any. Nevertheless, this saved energy is used afterwards (for $t>6 \cdot 10^5$), allowing the \emph{Heuristic} to deploy some applications that the \emph{Mixed} strategy cannot. Finally, coming back to Fig.~\ref{fig:ACONApps}, it was shown that the \emph{Heuristic} has a near-flat behavior with the moving cost while for the \emph{Max-min} and \emph{Mixed} strategies the performance degrades as the moving cost grows. This can be better explained now: The \emph{Heuristic} performs a low number of movements and some applications are rejected because of this. As the moving cost rises,  the \emph{Heuristic} saves more energy with respect to the other strategies by rejecting those applications. Thus, later on, more applications consuming few resources that can no longer be admitted by the \emph{Max-min} or \emph{Mixed} strategies can still be deployed by the \emph{Heuristic}.

\begin{center}
\begin{table*} [!b]
\renewcommand{\arraystretch}{1.3}
\centering
\caption{Scenario topologies.}
\begin{small}
	\centering
	 \adjustbox{max width=\textwidth}{
	\begin{tabular}{| r || c | c | c | c | c | c |}
 		\hline
 		\multirow{2}{1cm}{\textbf{}} & \multicolumn{6}{c|}{\textbf{Scenario}} \\ \cline{2-7}
& 1 &  2 & 3 & 4 & 5 & 6\\ \hline \hline
		Size & $100 \times 100$ m & $141 \times 141$ m & $173 \times 173$ m & $200 \times 200$ m & $224 \times 224$ m & $245 \times 245$ m\\ 
	Number of nodes& 18 & 36 & 54 &72 &90 &108\\ 
	Number of sinks& 1& 2& 3 &  4&  5 &  6\\ 
	Number of applications& 100 &  200 &  300 &  400  &  500  &  600\\ 
	Arrival rate & $0.5$ app/hour & $1$ app/hour & $1.5$ app/hour & $2$ app/hour &$2.5$ app/hour &$3$ app/hour  \\
		
				\hline
   	\end{tabular}}
\end{small}
\label{table:scenaries}
\end{table*}
\end{center} 

\section{Conclusion}
	\label{sec:conclusions}
In this work, we have studied the problem of virtual sensor network management from the perspective of a shared sensor network infrastructure provider that leases its physical resources to multiple concurrent applications/application providers. Namely, a joint optimization framework has been introduced to solve the problem of Application/service Admission Control and wireless Sensor Network Slicing (AAC-SNS). The proposed framework optimally decides (i) if/when to admit new applications to the use of the physical infrastructure, (ii) how to allocate the physical resources of the shared infrastructure to the multiple concurrent applications, while considering constraints at the sensor node level  (processing power and storage) as well as at the network level (available bandwidth, shared communication technologies, routing). The results of the proposed optimization framework have been compared against the ideal scenario where the shared sensor network infrastructure provider has full knowledge of the application/service request across time, which is introduced as a performance benchmark. Moreover, a greedy heuristic is also proposed to obtain close-to-optimal solutions of joint AAC-SNS problem in short computation time.

% or


\section*{Acknowledgment}
This work has been supported by the Spanish Government through the grant TEC2014-52969-R from the Ministerio de Ciencia e Innovaci\'on (MICINN), Gobierno de Arag\'on (research group T98), the European Social Fund (ESF) and Centro Universitario de la Defensa through project CUD2016-17.

\ifCLASSOPTIONcaptionsoff
  \newpage
\fi



\bibliographystyle{IEEEtran}
\bibliography{mybibfile}
%\begin{thebibliography}{1}



%\newpage


%\vfill


\end{document}